\documentclass[structabstract]{aa}

\usepackage{graphicx}
\usepackage{amsmath}
\usepackage{txfonts}
\usepackage{natbib}
\bibpunct{(}{)}{;}{a}{}{,}

\begin{document}

   \title{Regular Oscillation Sub-spectrum of Rapidly Rotating Stars}

   \subtitle{}

   \author{M. Pasek  \inst{\ref{inst1},\ref{inst2},\ref{inst3},\ref{inst4}}\fnmsep\thanks{\email{pasek@irsamc.ups-tlse.fr}}
          \and F. Ligni\`eres \inst{\ref{inst1},\ref{inst2}}
	  \and B. Georgeot \inst{\ref{inst3},\ref{inst4}}
	  \and D.R. Reese \inst{\ref{inst5}}
          }

   \institute{CNRS; IRAP; 14, avenue Edouard Belin, F-31400 Toulouse, France\label{inst1}
		\and Universit\'e de Toulouse; UPS-OMP; IRAP; Toulouse, France \label{inst2}
		\and CNRS; LPT (IRSAMC); F-31062 Toulouse, France \label{inst3}
		\and Universit\'e de Toulouse; UPS; Laboratoire de Physique Th\'eorique (IRSAMC); F-31062 Toulouse, France \label{inst4}
		\and Institut d'Astrophysique et G\'eophysique de l'Universit\'e de Li\`ege, All\'ee du 6 Ao\^ut 17, 4000 Li\`ege, Belgium \label{inst5}
	     }

   \date{Received May 30, 2012/ Accepted ??}

 
  \abstract
   {}
   {We present an asymptotic theory that describes regular frequency spacings of pressure modes in rapidly rotating stars.}
   {We use an asymptotic method based on an approximate solution of the pressure wave equation constructed from a stable periodic solution of the ray limit. 
   The approximate solution has a Gaussian envelope around the stable ray, and its quantization yields the frequency spectrum.}
   {We construct semi-analytical formulas for regular frequency spacings and mode spatial distributions of a subclass of pressure modes in rapidly rotating stars.
   The results of these formulas are in good agreement with numerical data for oscillations in polytropic stellar models. 
   The regular frequency spacings depend explicitly on internal properties of the star, and their computation for different rotation rates gives new insights on the evolution of mode frequencies with rotation.}
   {}

   \keywords{Asteroseismology - Chaos -  Methods: analytical - Stars: oscillations - Stars: rotation - Waves}

   \maketitle
%

\section{Introduction}
The field of asteroseismology has now reached its age of maturity with the exploitation of space missions CoRoT \citep{corotteam} and Kepler \citep{keplerteam} that are gathering stellar light curves with high accuracy. However, there are still unresolved issues that hinder the successful pairing of light curve frequencies with pulsation modes, which is crucial to obtain detailed information on the inner structure of observed stars. One of these issues is the rapid rotation of a star around its axis, since the exact nature of rotational effects on pulsation modes is not known. In particular, the centrifugal flattening (e.g. \citet{monnier}) affects the spectrum of pressure modes (p-modes) in a complex way \citep{lg2009}. This difficulty mainly concerns non-evolved massive and intermediate-mass pulsating stars which are typically rapid rotators \citep{royer}. Recently though, hints of regular frequency spacings have been found in
the spectrum of rapidly rotating $\delta$ Scuti stars observed with CoRoT \citep{garciaher2009, manteg2012}, and this could ease future mode identification.

\begin{figure}
   \centering
   \includegraphics[width=\linewidth]{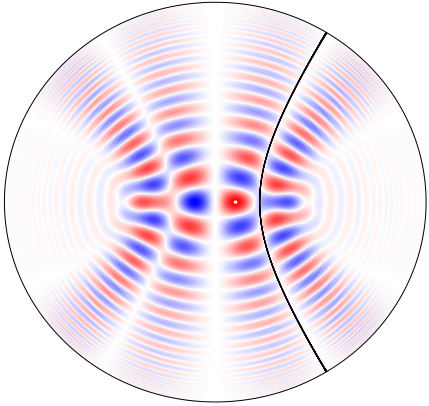}
   \caption{(Colour online) Pressure amplitude $P\sqrt{d/\rho_0}$ on a meridian plane for a polytropic stellar model, with $d$ the distance to the rotation axis and $\rho_0$ the equilibrium density. The mode shown corresponds to $n=50$, $\ell=1$ and $m=1$ at a rotation rate of $\Omega/\Omega_K =0.300$, where $\Omega_K=(GM/R_{eq}^3)^{1/2}$ is the limiting rotation rate for which the centrifugal acceleration equals the gravity at the equator, $M$ being the stellar mass and $R_{eq}$ the equatorial radius. Colors/grayness denote pressure amplitude, from red/gray (maximum positive value) to blue/black (minimum negative value) through white (null value). The thick black line is the ray $\gamma$ located in the center of the main stable island.}
\label{modefig}
\end{figure}

The recent development of accurate numerical models has enabled progress in the comprehension of pulsation modes in rapidly rotating stars. It has been found in particular \citep{letal2006,retal2008,reese2009} that in the rapidly rotating regime a subset of p-modes shows approximate regular frequency spacings in the form:
\begin{equation}
\label{regulr}
\omega_{n,\ell,m} \simeq \Delta_n n + \Delta_\ell \ell + \Delta_m |m| + \alpha,
\end{equation}
where frequencies $\omega_{n,\ell,m}$ are given in the corotating frame. Quantum numbers $n$, $\ell$ and $m$ correspond to node numbers of the mode amplitude distributions, $\Delta_n$, $\Delta_\ell$ and $\Delta_m$ are frequency regularities, and $\alpha$ is a constant term. The approximate formula in Eq. (\ref{regulr}) shows a better agreement with numerical results towards high-frequencies, thus suggesting that this relation is of an asymptotic nature. It should also be noted that, from computations of disk-averaging factors, the p-modes following Eq. (\ref{regulr}) are expected to be among the most visible ones \citep{lg2009}. An example of such a mode can be seen in Fig. \ref{modefig}. 

The frequency spacings of Eq. (\ref{regulr}) are notably similar to the regularities described by Tassoul's asymptotic formula \citep{tassoul80} for low degree p-modes in non-rotating stars. Tassoul's formula at leading order is 
\begin{equation}
\label{tassoul}
 \omega_{n,\ell} \simeq \Delta \left(n_s + \frac{\ell_s}{2} + \frac{1}{4} + \alpha_s \right)~, 
\end{equation}
with the large frequency separation
\begin{equation}
\label{largsep}
 \Delta = 2 \pi \left(2 \int^R _0 \frac{d r}{c(r)} \right)^{-1} ~,
\end{equation}
where $c(r)$ is the radially inhomogeneous sound speed, and $R$ the stellar radius. The integer $n_s$ is the node number of the radial component of the mode, while $\ell_s$ is the degree of the associated spherical harmonics, and $\alpha_s$ depends on surface properties. Tassoul's theory has proved to be very useful for interpretating solar-like oscillations in slowly rotating stars. Indeed, the formula relates observable quantities, such as the regular frequency spacing $\Delta$, to physical properties of stellar interiors. For rapidly rotating stars, it would be clearly desirable to gain insights on the underlying physics of the potentially observable regular spacings $\Delta_n$, $\Delta_\ell$ and $\Delta_m$ by a similar asymptotic analysis. In this paper we derive a formula for these regular frequency spacings in the asymptotic regime.

The generalization of the p-mode asymptotic theory to rapidly rotating stars is not trivial. Tassoul's theory requires separation of variables, which is no longer possible when the star is flattened by rotation. For non-separable wave systems, a well-known technique to obtain eigenmodes is to study the short-wavelength limit of the propagating waves. This limit gives an equation for the propagation of rays that is similar to the geometrical optics limit of electromagnetism, or the classical limit in quantum mechanics. Then, by imposing quantization conditions on the phase of waves propagating on these rays, one obtains the eigenmodes of the wave system. This technique was first developed in the context of quantum physics, and is often called semiclassical quantization. 

For spherical stars, the ray limit of pressure waves has been previously used to recover the Tassoul asymptotic formula from the Einstein-Brillouin-Keller (EBK) quantization of ray dynamics \citep{gough}. This analytical approach is possible only when the ray system is integrable. A dynamical system is said to be integrable when it has as many conserved quantities (energy, angular momentum, etc.) as degrees of freedom \citep{ott}. In rapidly rotating stars, there are not enough conserved quantities to ensure integrability of the ray dynamics. Indeed, in \citet{lg2008, lg2009}, it has been found that acoustic rays in rotating stars have a very different dynamical behavior depending on their initial conditions in position-momentum space (the so-called phase space). 
For a polytropic stellar model, the numerical integration of the equations for acoustic rays displayed various types of solutions. Indeed, one can obtain either stable rays staying on torus-shaped surfaces in phase space which form structures such as stable islands, or chaotic rays that are dense and ergodic on a phase space volume \citep{ott}. 

A similar behavior has been found in many systems studied in the field of theoretical physics known as quantum chaos or wave chaos \citep{gutz}. This field has among its objectives to analyze quantum (resp. wave) systems whose classical (resp. short-wavelength) limit is partly or fully chaotic. In this framework, one can predict the existence of some eigenfunctions (resp. mode amplitudes) and energies (resp. frequencies) of the quantum (resp. wave) system from the different structures that are present in the classical system phase space \citep{perci,berob}. In the stellar pulsation setting, \citet{lg2008, lg2009} found that the mixed (i.e. regular and chaotic) character of the acoustic ray dynamics in rapidly rotating stars results in a classification of p-modes in two broad families: regular modes either associated with stable islands or whispering gallery zones, and chaotic modes associated with ergodic regions in phase space. 
For the regular modes associated with stable islands, the so-called island modes, it is known to be possible to 
obtain approximate analytical solutions by solving the wave equation in the vicinity of a periodic stable ray \citep{bb}. A simple application of such a method is found in modes of optical resonators, where the periodic stable light ray is a straight line between two reflecting mirrors \citep{kogelnik66}. These methods have been previously employed to obtain modes of more complex lasing \citep{tureci} and electronic \citep{zalip} cavities as well as quantum chaos systems \citep{vagov}. In this paper, we apply this approach to rapidly rotating stars. 

In the present analysis, we thus construct an asymptotic formula for regularities in the p-mode spectrum of rapidly rotating stars. Part of the results were already presented in the short communication of \citet{pasek}. In the present paper we give a detailed derivation of these results, specify their domain of validity,  extend them with a study of rotational splittings, and explore their astrophysical applications.

The paper is organized as follows. In Sect. \ref{pmodes} we present the wave equation for p-modes in rotating stars and its asymptotic limit leading to an equation for acoustic rays. 
In Sect. \ref{regmodes} we use a stable periodic solution of the ray dynamics to obtain a semi-analytical formula for the associated p-modes, and to derive a formula for the associated regular frequency spacings. 
We then compare the results obtained from the derived formulas for mode frequencies and spatial distributions with numerical results (Sect. \ref{compnum}). 
Finally, we suggest directions on how these results could be used for the asteroseismic diagnosis of rapidly rotating stars by discussing the phenomenological implications of the theory in Sect. \ref{pheno}. 


\section{P-modes in rotating stars and their asymptotic limit}
\label{pmodes}

In Sect. \ref{pmodeseq}  we introduce the wave equation for p-modes in rotating stars. We then present the asymptotic limit of this equation in order to obtain an equation for the dynamics of acoustic rays (Sect. \ref{pmodesas}).
\subsection{Pressure modes in rotating stars}
\label{pmodeseq}
We start with the equation for small adiabatic time-harmonic perturbations of the pressure field in a self-gravitating gas. Since we are interested in obtaining an asymptotic theory for p-modes in the high-frequency regime, we use the Cowling approximation (i.e. we neglect the perturbations of the gravitational potential), an approximation known to be valid for high-frequency perturbations in non-rotating stars \citep{zebook}. We also neglect the Coriolis force. Indeed, in the high-frequency regime, the time scale associated with this force is much longer than the mode period, and thus the influence of the Coriolis force on pulsation frequencies is weak. This has been numerically checked in \citet{letal2006,retal2006,retal2008}. In the asymptotic regime of p-modes, the oscillation frequencies are far greater than the Brunt-V\"ais\"al\"a frequency and thus we can discard the terms corresponding to gravity waves. With these assumptions, the equation for pressure perturbations is a Helmholtz equation such that
\begin{equation}
\label{helm}
 \Delta \Psi + \frac{\omega ^2 - \omega_c ^2}{c_s ^2} \Psi = 0~,
\end{equation}
where $\Psi=\hat{P}/f$ is the complex amplitude associated with the pressure perturbation $P = \mathrm{Re}[\hat{P}\exp(- i \omega t )]$, $f$ is a function of the background model, $\omega_c$ is the cut-off frequency of the model and $c_s$ its inhomogeneous sound velocity \citep[for a detailed derivation of this equation see][]{lg2009}. The stellar model is not spherically symmetric due to the centrifugal distortion, but is however cylindrically symmetric with respect to the rotation axis. Therefore, we can write the pressure field as $\Psi = \Psi_m \exp ( i m \phi)$ where $m$ is an integer and $\phi$ is the azimuth angle of spherical coordinates. By inserting this expression in Eq. (\ref{helm}) we obtain (cf. Sect. \ref{2D})
\begin{equation}
\label{helm2}
 \Delta \Phi_m + \frac{1}{c_s ^2}\left( \omega^2 - \omega_c ^2 - \frac{c_s ^2 \left( m^2 - \frac{1}{4} \right)}{d^2} \right) \Phi_m = 0~,
\end{equation}
where $d$ is the distance to the rotation axis. The new mode amplitude $\Phi_m$ is such that $\Phi_m =\sqrt{d} \Psi_m $.
We introduce a renormalized sound velocity:
\begin{equation}
 \tilde{c}_s = \frac{c_s}{\sqrt{1-\frac{1}{\omega^2} \left( \omega_c^2+ \frac{c_s^2\left( m^2 - \frac{1}{4}\right)}{d^2}\right)}}~.
\end{equation}
We notice that besides its spatial dependence, $\tilde{c}_s$ also depends on $\omega$ and $m$, and that $m$ is taken as a parameter for the two-dimensional wave equation Eq. (\ref{helm2}). 

\subsection{Ray limit of p-modes}
\label{pmodesas}
In non-rotating stars, the asymptotic theory of high frequency p-modes has first been derived by \citet{vandakurov1967}, and \citet{tassoul80}. The method was to use the spherical symmetry of the star to reduce the problem to a one-dimensional equation in order to obtain the mode frequencies. This method is not applicable when the centrifugal force breaks the spherical symmetry of the star. In this case though, one can study the short-wavelength limit ($\omega\rightarrow \infty$) of the wave equation Eq. (\ref{helm}) (as detailed in \citet{lg2009}). This provides a Hamiltonian system describing the propagation of acoustic rays. The Hamiltonian has been derived in \citet{lg2009} as
\begin{equation}
 H = - \frac{\vec{\tilde{k}_p}^2}{2} + \frac{1}{2 c_s ^2} \left(1-\frac{\omega_c ^2}{\omega ^2} - \frac{c_s ^2 m ^2}{\omega ^2 d ^2} \right)~,
\label{wkb}
\end{equation}
where the frequency-scaled wavevector $\vec{\tilde{k}_p}$ is the projection of $\vec{\tilde{k}}=\vec{k}/\omega$ onto the meridional plane of the star. We notice that this expression has been derived from the short-wavelength limit of the three dimensional wave equation Eq. (\ref{helm}) and then, projected onto the corotating meridian plane. An alternative derivation would be to start from the two-dimensional wave equation Eq. (\ref{helm2}). In this case, the ray limit yields the same expression with the addition of the $1/4$ factor of Eq. (\ref{helm2}) that accounts for the impossibility of acoustic rays to go through the rotation axis (i.e. $d=0$). Throughout the paper we use Eq. (\ref{wkb}) as the Hamiltonian for acoustic rays. 

\begin{figure}
\centering
\includegraphics[width=\linewidth]{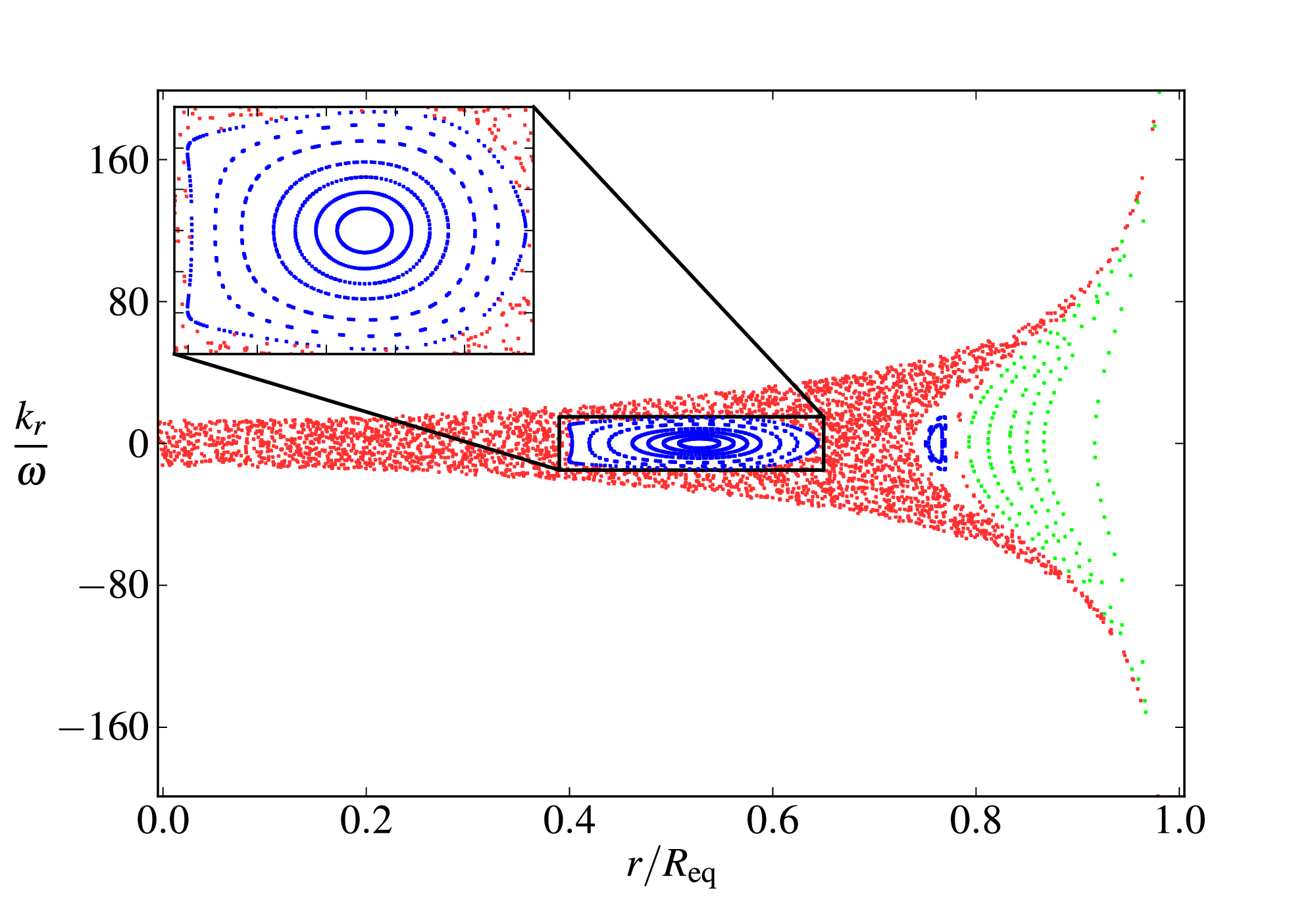}
\caption{(Colour online) Poincar\'e Surface of Section (PSS) at the rotation rate $\Omega/\Omega_K=0.589$ for quantum number $m=0$. Each dot corresponds to the crossing of an acoustic ray with the equatorial half-plane in the $(r/R_{eq},k_r/\omega)$ phase space, $r$ being the radial coordinate and $k_r$ the associated momentum. $R_{eq}$ is the equatorial radius and $\omega$ the mode frequency. Red/dark gray denotes a chaotic ray, green/light gray a whispering gallery ray, blue/black a stable island ray (see text). Upper inset is a close-up of the main stable island.
}
   \label{pss}
\end{figure}

To probe the integrability property of a dynamical system, it is convenient to use the Poincar\'e surface of section (PSS), a standard tool in dynamical systems theory \citep{gutz,ott} to visualize the structures in phase space. A PSS is a lower dimensional slice of phase space. The acoustic ray dynamical system in the meridional plane has two degrees of freedom, which gives a phase space of dimension four (two for positions, and two for momenta). There is one conserved quantity in the form of the acoustic wave frequency, so the dynamics belongs to a three-dimensional manifold in phase space. By fixing an additional position or momentum coordinate, we obtain a two-dimensional PSS which can be easily visualized. An example of such a section for our system is shown in Fig. \ref{pss}. Different choices of PSS variables are possible, some of which are presented in \citet{lg2009}. We have here chosen to fix the colatitude $\theta=\pi/2$, so that the PSS corresponds to the crossing of rays with the equatorial half-plane. 
We thus display a section in coordinates $(r/R_{eq},k_r/\omega)$ where $k_r$ is the norm of the radial wavevector, $\omega$ the mode frequency, and $R_{eq}$ the equatorial radius (that may be greater than the polar radius since the star is flattened by rotation). 
In such plots, each dot corresponds to the crossing of an acoustic ray with the PSS. Successive dots from a single ray will form lines in integrable zones, or fill surfaces densely in chaotic zones. We see in Fig. \ref{pss} that when the rotation rate $\Omega/\Omega_K$ (where $\Omega_K=(GM/R_{eq}^3)^{1/2}$ is the limiting rotation rate) is large, different structures coexist in the system phase space: stable islands correspond to concentric circles around a stable ray, whispering gallery rays to lines near the surface, and chaotic zones to densely filled areas \citep[for more details, see][]{lg2009}. In this paper, we will focus on the 2-periodic stable island which is the main stable island (shown in the inset of Fig. \ref{pss}). The rays' dynamics is very sensitive to the rotation rate, so the PSS will be different for each rotational velocity. Indeed, the locus in phase space of the main stable island changes as rotation increases. The major change happens when the central ray of the 2-periodic stable island undergoes a bifurcation at $\Omega/\Omega_K\simeq0.26$. 
For slow rotation rates, the central ray of the island is 
located on the polar axis, and through this bifurcation it transforms into two stable rays surrounding one unstable ray on the polar axis. Then, as rotation increases, the stable island will coast away from the polar axis. This bifurcation will be of some importance in the following, when we will show that one can construct approximate eigenmodes of the wave system from this 2-periodic stable island.  In Fig. \ref{modefig}, one can see an example of an island mode obtained from a full-numerical computation, together with the central ray of the 2-periodic stable island for the same rotation rate and quantum number $m$. 


\section{Semi-analytical method for island modes}
\label{regmodes}
In this section we construct an asymptotic approximation of a subset of regular p-modes associated with a stable periodic ray. 
The method is based on the works of Babich and coworkers (see \citet{bb} and references therein for the general formalism, and e.g. \citet{zalip, vagov} for some applications).  
It consists in deriving an approximation of the wave equation in the vicinity of the ray (Sect. \ref{harmeq}), finding Gaussian wavepacket solutions related to the stability properties of the ray (Sect. \ref{hermigau}), and then deriving the asymptotic formula for the frequencies from a quantization condition  (Sect. \ref{tass}).

\subsection{Approximate wave equation in the vicinity of a stable ray}
\label{harmeq}
For a given rotation rate and quantum number $m$, we start with the central periodic ray of the main stable island (see Sect. \ref{pmodes}). This ray must be computed by numerically evaluating the Hamiltonian equations derived from Eq. (\ref{wkb}). In the following we will call this ray $\gamma$. The first step is to write the wave equation Eq. (\ref{helm2}) in the vicinity of $\gamma$. For this, we use a local orthonormal coordinate system $(s,\xi)$ defined as $\vec{r}=s\vec{T} + \xi \vec{N}$ where $s$ is the arc length along the ray, $\vec{T}$ the unit tangent vector, $\xi$ the transverse coordinate and $\vec{N}$ the unit vector normal to $\vec{T}$. The two basis vectors are related by the curvature $\kappa(s)$ of the ray as follows:
\begin{equation}
 \kappa (s) = - \frac{d \vec{N}}{d s} \cdot \vec{T}~.
\end{equation}
In this coordinate system, the wave equation Eq. (\ref{helm2}) reads
\begin{equation}
\label{helm3}
 \frac{1}{h_s h_\xi} \left( \frac{\partial}{\partial s} \left( \frac{h_\xi}{h_s} \frac{\partial \Phi _m}{\partial s}\right) + \frac{\partial}{\partial \xi} \left( \frac{h_s}{h_\xi} \frac{\partial \Phi _m}{\partial \xi}\right) \right) + \frac{\omega^2}{\tilde{c}_s(s,\xi) ^2} \Phi _m = 0~,
\end{equation}
where the scale factors $h_s$ and $h_\xi$ \citep{arf} are
\begin{equation}
 h_s ^2 = \left( \vec{T} - \xi \kappa (s) \vec{T} \right) ^2 = ( 1 - \xi \kappa (s) ) ^2
\end{equation}
and
\begin{equation}
 h_\xi ^2 = \vec{N}(s) ^2 = 1~.
\end{equation}
In the vicinity of the ray $\gamma$, that is for small $\xi$, the terms of Eq. (\ref{helm3}) are simplified using:
\begin{equation}
 \frac{1}{( 1 - \xi \kappa (s) )} \sim 1 + \xi \kappa (s) + \xi ^2 \kappa (s) ^2 + O(\xi^3)~,
\end{equation}
\begin{equation}
 \frac{1}{( 1 - \xi \kappa (s) ) ^2} \sim 1 + 2 \xi \kappa (s) + 3 \xi ^2 \kappa (s) ^2 + O(\xi^3)~,
\end{equation}
\begin{equation}
 \frac{1}{( 1 - \xi \kappa (s) ) ^3} \sim 1 + 3 \xi \kappa (s) + 6 \xi ^2 \kappa (s) ^2 + O(\xi^3)~,
\end{equation}
and
\begin{equation}
 \frac{1}{\tilde{c}_s (s,\xi) ^2} = \frac{1}{\tilde{c}_s (s,0) ^2} + \frac{\partial \left(1/\tilde{c}_s (s,\xi) ^2 \right)}{\partial \xi }\Big |_{\xi=0} \xi + O(\xi^2)~.
\end{equation} 
We then express the function $\Phi_m (s,\xi)$ in terms of a WKB ansatz as
\begin{equation}
\label{ansatz}
\Phi_m (s,\xi)=\exp(i\omega \tau)U_m(s,\xi,\omega)~,
\end{equation}
where $\tau$ is an unknown function of position. The fundamental assumption underlying the theory of Babich is that, as $\omega \rightarrow +\infty$, the mode is localized on the acoustic ray and that its transverse extent scales as $1/\sqrt{\omega}$. Such a solution can be found by assuming that the transverse variable $\xi$ scales as
\begin{equation}
 \xi=O\left(1/\sqrt{\omega}\right)~.
\end{equation}
Then, from an expansion of Eq. (\ref {helm3}) in $\omega$, one obtains at the dominant order that the WKB phase in Eq. (\ref{ansatz}) depends only on $s$ as $d\tau=ds/\tilde{c}_s$. At the next order in $\omega$ one finds a parabolic equation for the function $V_m$:
 \begin{equation}
\label{parab}
 \frac{\partial ^2 V_m}{\partial \nu ^2} - K(s) \nu ^2 V_m + \frac{ 2 i}{\tilde{c}_s (s)} \frac{\partial V_m}{\partial s}  = 0 ~,
 \end{equation}
with 
\begin{equation}
K(s) = \frac{1}{\tilde{c}_s(s)^3} \frac{\partial ^2 \tilde{c}_s}{\partial \xi ^2}\Big |_{\xi=0}~,
\end{equation}
where we introduced the scaled coordinate $\nu=\sqrt{\omega}~\xi$ and $V_m=U_m / \sqrt{\tilde{c}_s}$. 


\subsection{Solutions of the parabolic wave equation}
\label{hermigau}
To find a solution to Eq. (\ref{parab}), we first find a solution at a fixed arc length $s$, and then study how this solution must evolve with $s$. At fixed $s$, the first two terms of Eq. (\ref{parab}) correspond to the equation for a quantum harmonic oscillator in the direction $\vec{N}$ transverse to $\gamma$. Thus, as we know from quantum mechanics \citep{cota}, a solution of this equation is a Gaussian wavepacket, transverse to the ray, that we write as
 \begin{equation}
\label{ground}
  V_{m}^0 = A (s) \exp \left( i \frac{\Gamma(s)}{2} \nu ^2 \right) ~,
 \end{equation}
with $\Gamma$ an unknown complex-valued function. To find the variation of this Gaussian wavepacket along the ray $\gamma$, we introduce a solution of this form in the parabolic equation Eq. (\ref{parab}) and obtain a Riccati equation for $\Gamma$
 \begin{equation}
\label{ricca}
  \frac{1}{\tilde{c}_s} \frac{d \Gamma}{d s} + \Gamma ^2 + K = 0~,
 \end{equation}
 and a simple form for the factor $A$
 \begin{equation}
  \frac{1}{A}\frac{d A}{d s} = \frac{-\tilde{c}_s}{2} \Gamma ~.
 \end{equation}
In the following, we show that the equation for $\Gamma$ is related to the ray properties in the vicinity of $\gamma$ and can thus be solved from ray dynamics computations. First, using the variables $(z(s),p(s))$ defined as $\Gamma(s) = \frac{1}{z(s)}\frac{1}{\tilde{c}_s}\frac{d z (s)}{ d s}$ and $p(s)=\frac{1}{\tilde{c}_s}\frac{d z (s)}{ d s}$, Eq. (\ref{ricca}) is transformed into the Hamiltonian system:
\begin{align}
\label{ham1}
 \frac{d z}{d \tau}& = \tilde{c}_s^2 p \\
\label{ham2}
 \frac{d p}{d \tau}& =  - \tilde{c}_s^2 K z~,
\end{align}
where the (time-dependent) Hamiltonian function is
\begin{equation}
\label{norm}
H_0(p,z,\tau) = \tilde{c}_s^2 \frac{p^2}{2} + \tilde{c}_s^2 K \frac{z^2}{2}~,
\end{equation}
$\tau$ being the time coordinate. 
Equations (\ref{ham1}-\ref{ham2}) have two independent solutions $(z,p)$ and $(\bar{z},\bar{p})$ that are conjugate to each other. Of these two solutions, only one is physically relevant, i.e. corresponds to a localized wavepacket. According to Eq. (\ref{ground}) this happens if $\mathrm{Im}(\Gamma)>0$ for all $s$, and as shown in Sect. \ref{wro}, the sign of $\mathrm{Im}(\Gamma)$ stays constant along $\gamma$ since
\begin{equation}
\label{gamz}
 \mathrm{Im}(\Gamma)=\frac{1}{2}\frac{1}{|z|^2}~.
\end{equation}
The variation along $\gamma$ of the Gaussian wavepacket can now be linked to the dynamics of the acoustic rays nearby $\gamma$. Indeed, $\mathrm{Im}(\Gamma)$ has a simple expression in terms of the complex variable $z$, as shown in Eq. (\ref{gamz}). This variable, on the other hand, obeys Eqs. (\ref{ham1}-\ref{ham2}) which can be shown to be the same as the equation describing the deviation from $\gamma$ of a ray nearby $\gamma$ (see derivation in Sect. \ref{jaco}). The Hamiltonian in Eq. (\ref{norm}) is thus a local integrable approximation, also known as a normal form approximation \citep{arn}, to the full Hamiltonian for acoustic rays written in Eq. (\ref{wkb}).

Now, our task is to find the two linearly independent complex conjugate solutions of Eqs. (\ref{ham1}-\ref{ham2}). The terms in these equations depend only on quantities that are evaluated on the periodic ray $\gamma$. Therefore, these equations are periodic in $s$, or equivalently in $\tau$. Eqs. (\ref{ham1}-\ref{ham2}) can thus be written as:
\begin{equation}
 \label{ham3}
\frac{d}{d\tau} \begin{pmatrix} z \\ p \end{pmatrix} = \Sigma(\tau) \begin{pmatrix} z \\ p \end{pmatrix}~,
\end{equation}
where the matrix $\Sigma$
\begin{equation}
\label{ham4}
 \Sigma(\tau) = \begin{pmatrix} 0 & \tilde{c}_s^2 \\ -\tilde{c}_s^2 K & 0 \end{pmatrix}~,
\end{equation}
verifies $\Sigma(\tau+T_\gamma)=\Sigma(\tau)$, $T_\gamma=\oint_\gamma \frac{ds}{\tilde{c}_s}$ being the time period associated with $\gamma$. Then if $(z(\tau),p(\tau))$ is a solution of Eq. (\ref{ham3}), so is $(z(\tau+T_\gamma),p(\tau+T_\gamma))$ and the two solutions are related by the following linear map
\begin{equation}
\label{monod}
 \begin{bmatrix} z(\tau+T_\gamma)\\
p(\tau+T_\gamma)
\end{bmatrix}
= M 
\begin{bmatrix}z(\tau) \\
 p(\tau)
 \end{bmatrix}~,
\end{equation}
where $M$ is called the monodromy matrix (\citet{cvit}, and references therein). As $(z,p)$ describe ray deviations from $\gamma$, the matrix $M$ characterizes the stability of $\gamma$. 
As $\gamma$ is stable, we know that $|\mathrm{Tr}(M)| < 2$ and that the eigenvalues are of modulus one and complex conjugates of each other i.e. $\Lambda^\pm=\exp(\pm i \alpha)$ with $\alpha \in ]0,\pi [$ (cf. Sect. \ref{alg}), where $\alpha$ is called a Floquet phase or stability angle. Hence the two linearly independent solutions of Eq. (\ref{ham3}) can be written in the form: 
\begin{equation}
\label{floq}
 (z(\tau),p(\tau))^{\pm}=\exp\left(\pm i \frac{\alpha}{T_\gamma} \tau\right) u_{\pm}(\tau) \vec{v}^{\pm}~,
\end{equation}
where the functions $u_{\pm}(\tau)$ are periodic with period $T_\gamma$ and $\vec{v}^{\pm}$ are independent eigenvectors of the monodromy matrix $M$.

An expression of the monodromy matrix in terms of second derivatives of the action function $S$ can be derived \citep{bogo}. 
The action function $S$ is defined by a trajectory from the position $q_i$ to $q_f$ for a given energy or frequency $\omega$ \citep{gutz}. 
For our purposes, the action is written as
\begin{equation}
 S(q_i,q_f,\omega)=\int_{q_i}^{q_f} \frac{1}{\tilde{c}_s} d\sigma~,
\end{equation}
where $\sigma$ is the arclength along a ray nearby $\gamma$. 
If we write the monodromy matrix as
\begin{equation}
 \begin{pmatrix}
 z_f \\
 p_f  
 \end{pmatrix} = 
 \begin{pmatrix}
 M_{11}&M_{12} \\
 M_{21}&M_{22}
 \end{pmatrix}
 \begin{pmatrix}
 z_i \\
 p_i
 \end{pmatrix}~,
\end{equation}
we can express its components from the second derivatives of the action function $S$ with the following formulas
\begin{equation}
 \frac{\partial^2 S}{\partial z_i \partial z_f}= - \frac{1}{M_{12}} ~,~ \frac{\partial^2 S}{\partial z_i ^2} = \frac{M_{11}}{M_{12}}~,~\frac{\partial^2 S}{\partial z_f ^2} = \frac{M_{22}}{M_{12}}~,
\end{equation}
 where the positions $q_i$ and $q_f$ are written as $(s_i,z_i)$ and $(s_f,z_f)$, $z_i$ and $z_f$ being respectively the initial and final transverse positions of the neighboring ray after one period, and the derivatives are evaluated on the periodic ray. 
From these expressions, and the simple formula giving the roots of a second degree polynomial (cf. Sect. \ref{alg}), we can obtain an expression for the stability angle $\alpha$ as
\begin{equation}
 \label{eqalpha}
 \alpha = \arctan\left( \frac{\sqrt{-\mathrm{Tr}(M)^2+4}}{\mathrm{Tr}(M)} \right)~,
\end{equation}
where
\begin{equation}
 \mathrm{Tr}(M) = - \left(\frac{\partial^2 S}{\partial z_i \partial z_f} \right)^{-1} \left(\frac{\partial^2 S}{\partial z_i ^2}
 +\frac{\partial^2 S}{\partial z_f ^2} \right)~.
\end{equation}

We thus have obtained a solution of the approximate wave equation Eq. (\ref{parab}) in the form of a Gaussian wavepacket (Eq. (\ref{ground})), whose evolution along the ray $\gamma$ is given by Eqs. (\ref{ham3}-\ref{ham4}). 
It is possible to find other solutions of Eq. (\ref{parab}) that have a finite number of nodes in the direction transverse to $\gamma$. 
In the same way as in quantum mechanics \citep{cota}, these solutions can be obtained from Eq. (\ref{ground}) using the annihilation $\hat{a}$ and creation operator $\hat{a}^\dagger$ that are defined as
 \begin{equation}
\label{quop}
  \hat{a} = - i z \frac{\partial}{\partial \nu} - p \nu ~\mathrm{and}~ \hat{a} ^ \dagger = - i \bar{z} \frac{\partial}{\partial \nu} - \bar{p} \nu ~.
 \end{equation}
These operators have the commutation rule $\left[ \hat{a}, \hat{a} ^\dagger \right] = 1$, where the commutator is defined as $\left[ \hat{A}, \hat{B}  \right] = \hat{A}\hat{B}-\hat{B}\hat{A}$. 
It can be shown that the expression for the higher-order solutions is
 \begin{equation}
\label{recur}
V_m ^\ell =   (\hat{a} ^ \dagger ) ^ \ell V_m ^0~.
 \end{equation}
This defines a recurrence relation, whose solutions are proportional to Hermite-Gauss polynomials and can be written in the following way
\begin{equation}
\label{mode}
V_{m}^\ell (s,\nu) = \left( \frac{\bar{z}}{z} \right)^{\ell/2} H_\ell \left( \sqrt{\mathrm{Im}(\Gamma)}\nu \right) \frac{\exp \left( i \frac{\Gamma}{2} \nu ^2 \right)}{\sqrt{z}}~,
\end{equation}
with $H_\ell$ the Hermite polynomials  of order $\ell$. Finally, we can write the solutions of Eq. (\ref{helm2}) as
\begin{equation}
\label{modefull}
\Phi_{m}^\ell (s,\nu) =  \sqrt{\tilde{c}_s} V_{m}^\ell (s,\nu) \exp\left(i\omega\int \frac{ds}{\tilde{c}_s}\right)~.
\end{equation}


\subsection{Quantization condition and regular frequency spacings}
\label{tass}
\begin{figure}
   \centering
   \includegraphics[scale=0.7]{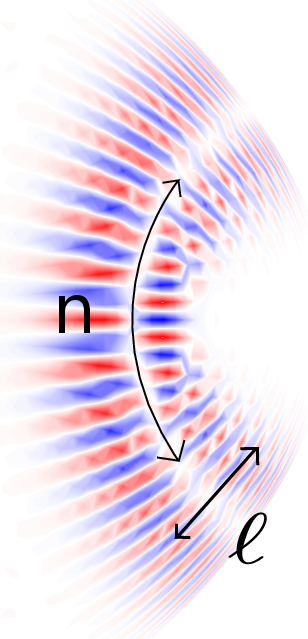}
   \hfill
   \includegraphics[scale=0.3]{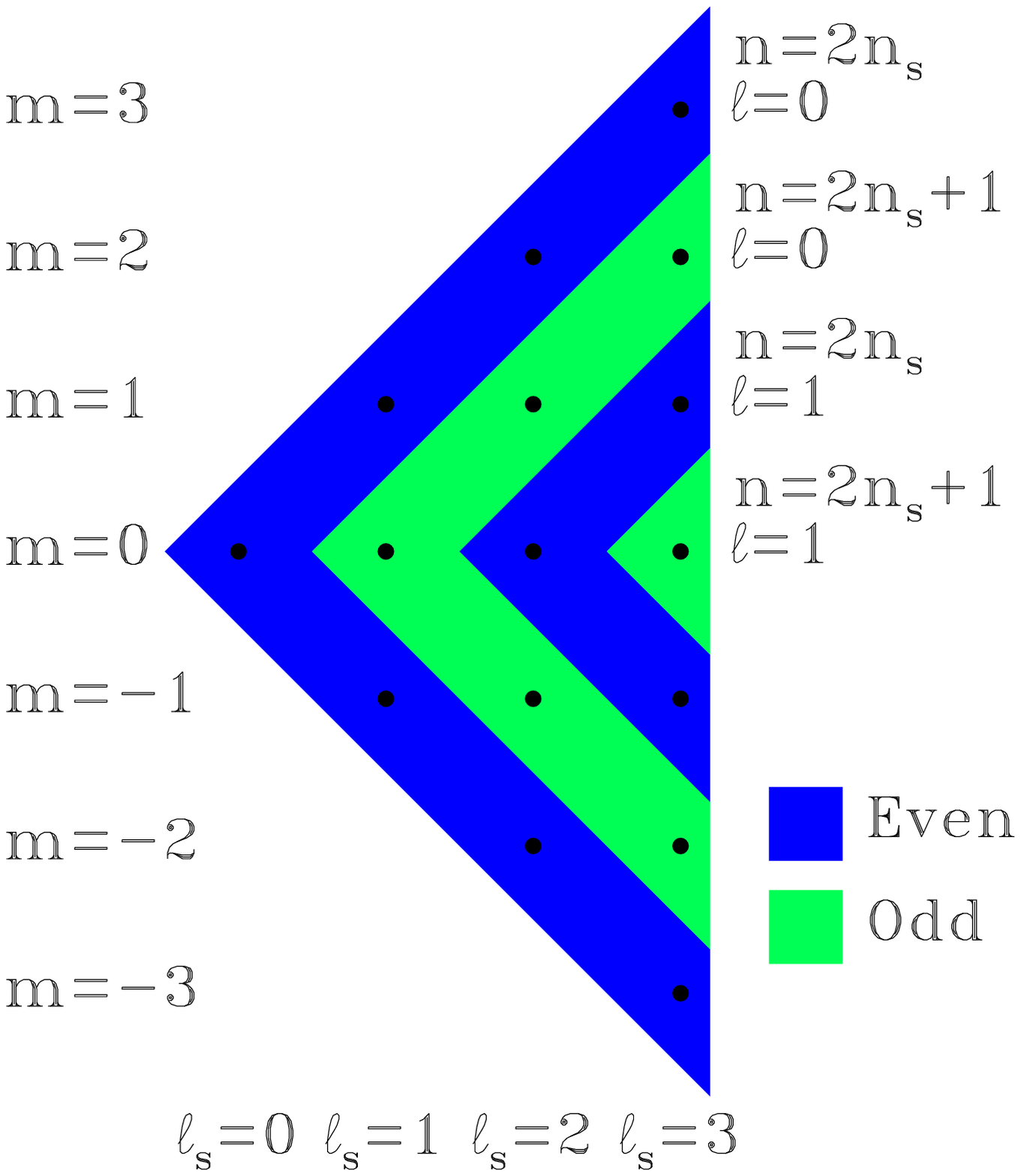}
   \caption{(Colour online) Left: Schematic representation of the mode labeling/quantum numbers used for island modes. Right: Illustration of the relation between the quantum numbers of spherical and island modes.  The quantum number $n$ is the number of nodes in the longitudinal direction of the island mode (along the ray $\gamma$), $\ell$ is the number of nodes in the transverse direction of the mode (transverse to $\gamma$) and $m$ is the number of azimuthal nodes. $n_s$, $\ell_s$ and $m_s$ are the quantum numbers of spherical modes. In the right figure, the multiplets of island modes correspond to diagonal colored bands, whereas the multiplets of spherical modes would have the form of vertical bands. The plotted mode on the left corresponds to $n=46$, $\ell=1$ and $m=0$.}
   \label{node}
\end{figure}

The quantization condition is based on the single-valuedness of the solution presented in Eq. (\ref{modefull}), and thus asserts that the phase accumulated by the function $\Phi_m^\ell$ over one period must be a multiple of $2\pi$. In the following, we assume without loss of generality that the eigenvalue which corresponds to the wavepacket localization is $\exp(+i\alpha)$. The contribution of function $V_m^\ell$ to the dynamical phase of $\Phi_m^\ell$ is obtained from Eq. (\ref{floq}) and Eq. (\ref{mode}). We obtain that the phase accumulated over one period is
\begin{equation}
\label{phase}
 \omega_{n,l,m} \oint_\gamma \frac{1}{\tilde{c}_s} ds   - \frac{\alpha+2\pi N_r}{2} - (\alpha+2\pi N_r) \ell = 2 \pi n + \pi~,
\end{equation}
where $T_\gamma=\oint_\gamma \frac{ds}{\tilde{c}_s}$ is the acoustic travel time along $\gamma$ and $\alpha$ the Floquet phase that is defined modulo $2\pi$. For our purposes, we must also take into account the number $N_r$ of multiples of $2\pi$ acquired by the phase. $N_r$ can be computed by following the evolution of the eigenvector $\vec{v}$ over one period, and we verified numerically that, alternatively, $N_r$ is also the winding number around $\gamma$ of a ray nearby $\gamma$ during one period. The last term in Eq. (\ref{phase}) is the Maslov phase \citep{gutz} that comes from the reflection of the wave on the boundaries. The formula for the frequencies, $\omega_{n,l,m}$, of island modes is thus
\begin{equation}
\label{regul}
  \omega_{n,l,m} = \frac{1}{\oint_\gamma \frac{ds}{\tilde{c}_s}} \left[2 \pi \left( n + \frac{1}{2} \right) + \left( \ell + \frac{1}{2} \right) \left( 2 \pi N_r + \alpha \right) \right]
 \end{equation}
or, in a form that makes the regularities more visible,
\begin{equation}
\label{reg1}
\omega_{n,l,m} = \delta_n(m) n + \delta_\ell(m) \ell + \beta(m)~,
\end{equation}
with the frequency regularities
\begin{equation}
\label{reg2}
  \delta_n = \frac{2 \pi}{\oint_\gamma \frac{ds}{\tilde{c}_s}} ~~ \mathrm{and} ~~ \delta_\ell = \frac{2 \pi N_r + \alpha}{\oint_\gamma \frac{ds}{\tilde{c}_s}}
 \end{equation}
and the constant term
\begin{equation}
\label{reg3}
\beta = \frac{\delta_n + \delta_\ell}{2}~.
 \end{equation}
The quantum numbers $n$ and $\ell$ correspond to node numbers of the p-mode, respectively in the longitudinal and transverse directions of the central ray $\gamma$ as illustrated in Fig. \ref{node}. This is to be contrasted with the case of a spherical mode where the most natural labeling are the quantum numbers of spherical harmonics. Eq. (\ref{regul}) is a semi-analytical formula since the quantities $T_\gamma$, $N_r$ and $\alpha$ must be computed numerically from the Runge-Kutta integration of the Hamiltonian equations for acoustic rays. The acoustic time, $T_\gamma$, is directly computed from $\gamma$ itself. From the intersections of a ray nearby $\gamma$ with the PSS (as can be seen in Fig. \ref{pss}), we compute the monodromy matrix $M$ that maps one intersection with the PSS to the next one. Then, by diagonalizing this matrix one obtains the Floquet phase $\alpha$ from its eigenvalues, and the functions $z$ and $\Gamma$ from its eigenvectors. It is important to note that for $m$ even, only modes 
symmetric with respect to the rotation axis exist. Since the preceding theory does not take this phenomenon into account, the theoretical value of $\delta_\ell$ is multiplied by two when the ray $\gamma$ coincides with the rotation axis, i.e. for rotation rates less than the bifurcation point.
Finally, it can be noted that a formula similar to Eq. (\ref{regul}) can also be obtained through the formalism of the Gutzwiller trace formula following the method of \citet{mil}. 


\section{Comparison with numerical results}
\label{compnum}
As the present asymptotic theory relies on various assumptions, its relevance for stellar seismology is not guaranteed and needs to be assessed through a comparison with exact calculations of realistic stellar models. 
In this section, the comparison is done with numerically computed modes in uniformly rotating polytropic models of stars.
The hypotheses of the asymptotic theory are the following : first, it is valid in the asymptotic regime, that is, for high enough frequencies.
Second, the island modes are constructed from a stable island of acoustic ray phase space. 
At null rotation such a structure does not exist, so we expect that the theory fails to describe spherical mode amplitudes and frequencies. 
A stable island immediately appears at non-zero rotation, but its phase space volume must be high enough for an island mode to exist. 
This volume increases with frequency and rotation (see \citet{lg2008,lg2009} for details).
Thus, for a given frequency range, the number of island modes starts from zero at small rotation rates and progressively increases
as the rotation and thus the phase space volume of the stable island grows.
Actually, low degree spherical modes become progressively island modes as rotation increases. 
Another assumption used in finding a solution to the wave equation is that the mode decays as $\propto 1/\sqrt{\omega}$ in the direction transverse to the periodic ray.
Finally, the theory also neglects the Coriolis force and the perturbations of the gravitational potential.

In the following, the asymptotic theory is compared with highly accurate computations of high frequency adiabatic modes in 
uniformly rotating polytropic stellar models with index $N=3$, the Coriolis effect and perturbations of the gravitational potential being taken into account.
The accuracy of these calculations,
described in detail in \citet{retal2006}, is very high (the relative precision on the frequencies is $10^{-7}$) and thus does not interfere with the present comparison. 
A large number of modes were followed from $\Omega/\Omega_K=0$ to $\Omega/\Omega_K=0.896$.
At zero rotation these modes are low degree $\ell_s \in \{0,1, 2,3\}$, high order $n_s \in [21,25]$ modes. At higher rotation rates,
they become island modes and can thus be labeled with $n$ and $\ell$, the number of nodes along and transverse to $\gamma$, respectively, as illustrated in Fig. \ref{node}.
The relation between the quantum numbers at zero and high rotation rates is \citep{r2008} :
\begin{align}
\label{qunum1}
 n& = 2 n_s + [(\ell_s+m_s) ~\mathrm{mod}~ 2]~,\\
\label{qunum2}
\ell& = \frac{\ell_s - |m_s| - [(\ell_s+m_s) ~\mathrm{mod}~ 2]}{2}~,\\
\label{qunum3}
m& = m_s~.
\end{align}
We remind here the reader that rotational multiplets are defined, in the non-rotating case, as a set of frequencies with identical $n_s$, $\ell_s$ quantum numbers but different values of $m_s$ for $m_s\in[-\ell_s,\ell_s]$. 
For rotating stars, we can define multiplets as frequencies with identical $n$ and $\ell$ but different $m\in\mathbb{Z}$, i.e. without any limiting value for $m$.
The relation between the two sets of quantum numbers and the two types of multiplets is illustrated in Fig. \ref{node}. In this figure the multiplets of island modes correspond to diagonal colored bands, whereas the multiplets at zero rotation would have the form of vertical bands.
We restricted ourselves to numerical modes with $\ell_s^\mathrm{max}=3$ so, in terms of island mode quantum numbers, the range of numerically computed modes is the one given in Table \ref{qunumtab}.
\begin{table}
\caption{Island mode quantum numbers $n$, $\ell$, $m$ of numerically computed modes, corresponding to $n_s \in [21,25]$, $\ell_s \in \{0,1,2,3\}$, $m_s\in[-\ell_s,\ell_s]$ in terms of spherical mode quantum numbers.}
\centering
\begin{tabular}{ccc}
\hline
\hline
$n$ & $\ell$ & $m$ \\
\hline
42, 44, 46, 48, 50 & 0 & -3, -2, \dots, 3 \\
42, 44, 46, 48, 50 & 1 & -1, 0, 1 \\
43, 45, 47, 49, 51 & 0 & -2, -1, \dots, 2 \\
43, 45, 47, 49, 51 & 1 & 0 \\
\hline
\end{tabular}
\label{qunumtab}
\end{table}
The associated numerical frequency spacings are defined as:
\begin{equation}
 \delta_n^N = \omega_{n+1,\ell,m}^N - \omega_{n,\ell,m}^N~,
\end{equation}
and
\begin{equation}
 \delta_\ell^N = \omega_{n,\ell+1,m}^N - \omega_{n,\ell,m}^N~.
\end{equation}
The semi-analytical asymptotic theory also requires determining the $\alpha$ term in Eq. (\ref{reg2}) numerically. To test the robustness of this calculation,
we checked that the frequency spacing $\delta_\ell$ only weakly depends on the choice of the ray nearby $\gamma$ that is used to compute $\alpha$. 
Also, the spacings $\delta_n$ and $\delta_\ell$ neither depend on the resolution of the background model nor on the integration parameters of the Runge-Kutta method. 


\subsection{Regular frequency spacings}
\label{compnumfreq}
\begin{figure}
   \centering
   \includegraphics[width=\linewidth]{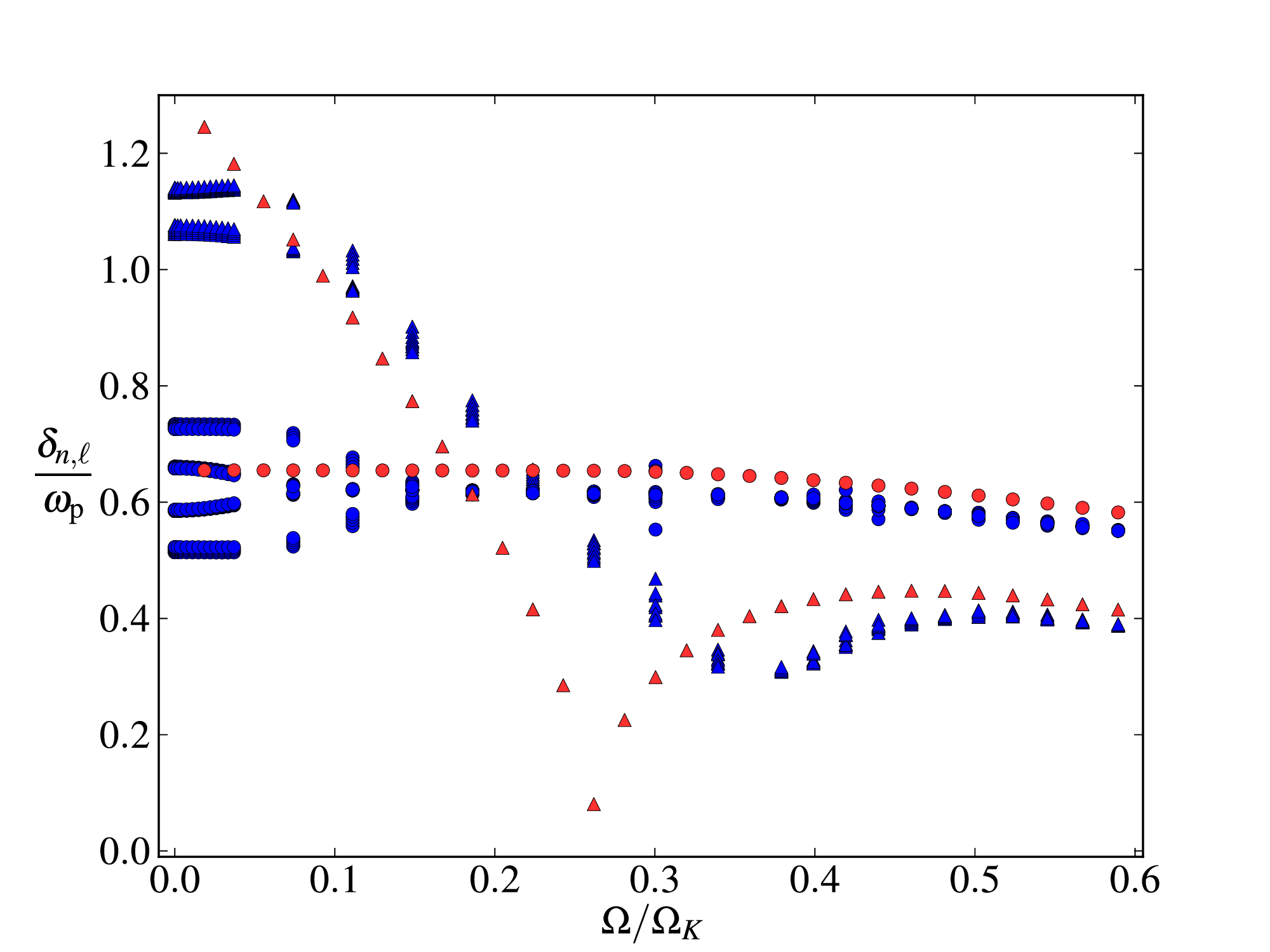}\\
   \includegraphics[width=\linewidth]{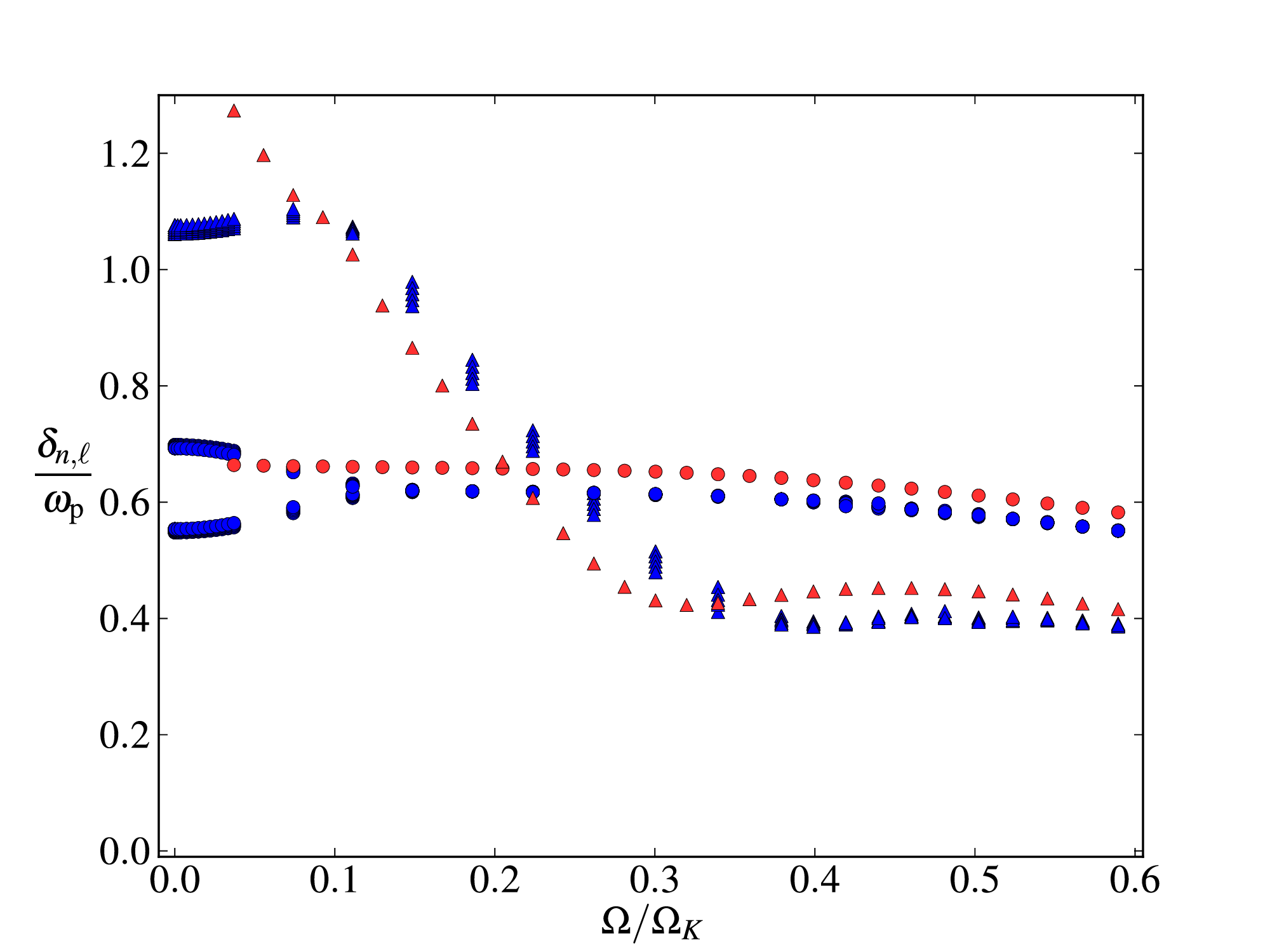}
   \caption{(Colour online) Comparison of frequency spacings $\delta_{n, \ell}$ between island modes, computed from numerical simulations and semi-analytical formulas for different values of $\Omega/\Omega_K$ (the frequency spacings are normalized by $\omega_p=\sqrt{GM/R_p^3}$ with $R_p$ the polar radius). Circles: $\delta_n$, triangles: $\delta_\ell$, red/dark gray: semi-analytical results, blue/black: numerical results. Numerical results correspond to different sets of $\delta_n^N$ and $\delta_\ell^N$ values, with $n\in [42,51]$, $\ell \in\{0, 1\}$ for $m=0$; and $n \in [42,51]$, $\ell=0$, $n \in\{42,44,46,48,50\}$, $\ell=1$ for $m\in\{-1,1\}$. Upper panel: $m=0$. Lower panel: $m=\pm 1$.}
   \label{regfig}
\end{figure}

According to Eq. (\ref{reg1}), the structure of the spectrum is characterized by the two spacings $\delta_n(m)$ and $\delta_\ell(m)$.
In Fig. \ref{regfig}, their semi-analytical and numerical values computed for $m=0$ and $|m|=1$ are compared as a function of the rotation rate.
One can see that the semi-analytical regularities $\delta_n$, $\delta_\ell$, and the full computations of high-frequency p-modes are in good agreement for almost all rotation rates. 
For $m=0$, around $\Omega/\Omega_K \simeq 0.26$, the agreement degrades significantly. 
In this rotation range, the ray $\gamma$ in the center of the main stable island undergoes a bifurcation from one stable ray on the polar axis to two stable rays surrounding one unstable ray. 
When such a bifurcation occurs, the eigenvalues of the monodromy matrix become $\Lambda^\pm = 1$, corresponding to a Floquet phase $\alpha=0~\mathrm{mod}~2\pi$ \citep[see e.g.][]{brack2001}. 
This is indeed what happens at $\Omega/\Omega_K \simeq 0.26$, as $\delta_\ell \propto \alpha$ goes to zero. 
Such a behavior conveys the non-validity of the present normal form approximation for rays undergoing a bifurcation.
One possibility would be to use other local approximations of the ray dynamics called uniform approximations \citep{schom}. 
The discrepancy coming from the bifurcation is not to be found for $m\neq0$, since the stable ray stays away from the polar axis and does not undergo a bifurcation as rotation increases. 

Although, as mentioned before, the theoretical and numerical frequency spacings are not expected to match for slow rotation rates, the discrepancies remain small in this rotation range. 
This is due to the fact that, as $\Omega/\Omega_K$ approaches zero, the stable ray is along the polar axis and, according to the expression of $\delta_n$, 
this implies that  $\delta_n = \Delta /2$, i.e. half the large separation  defined in Eq. (\ref{largsep}).
Now, using the first order of Tassoul's formula and the quantum numbers conversion rules Eqs. (\ref{qunum1}-\ref{qunum3}), it
is easy to see that, at zero rotation, $\delta_n^N= \omega^N_{n+1, \ell, m}-\omega^N_{n, \ell, m}$ is expected to be close to half the large separation.
Note also that the doubling of the numerical values observed in Fig. \ref{regfig} at small rotation rates is due to the small separation that appears at the next order of Tassoul's theory. 
Concerning $\delta_\ell$, the semi-analytical calculations indicates that $\delta_\ell$ goes to $2 \delta_n$, that is $\Delta$, for slow rotation rates. 
Again, Tassoul's formula applied to the $n$, $\ell$ quantum numbers shows that $\delta_\ell^N$ is close to $\Delta $. 
For these reasons, the frequency spacings $\delta_n$ and $\delta_\ell$ converge to the results of the first order of Tassoul's formula, though their derivation is formally not possible for non-rotating stars.

\begin{figure}
   \centering
   \includegraphics[width=\linewidth]{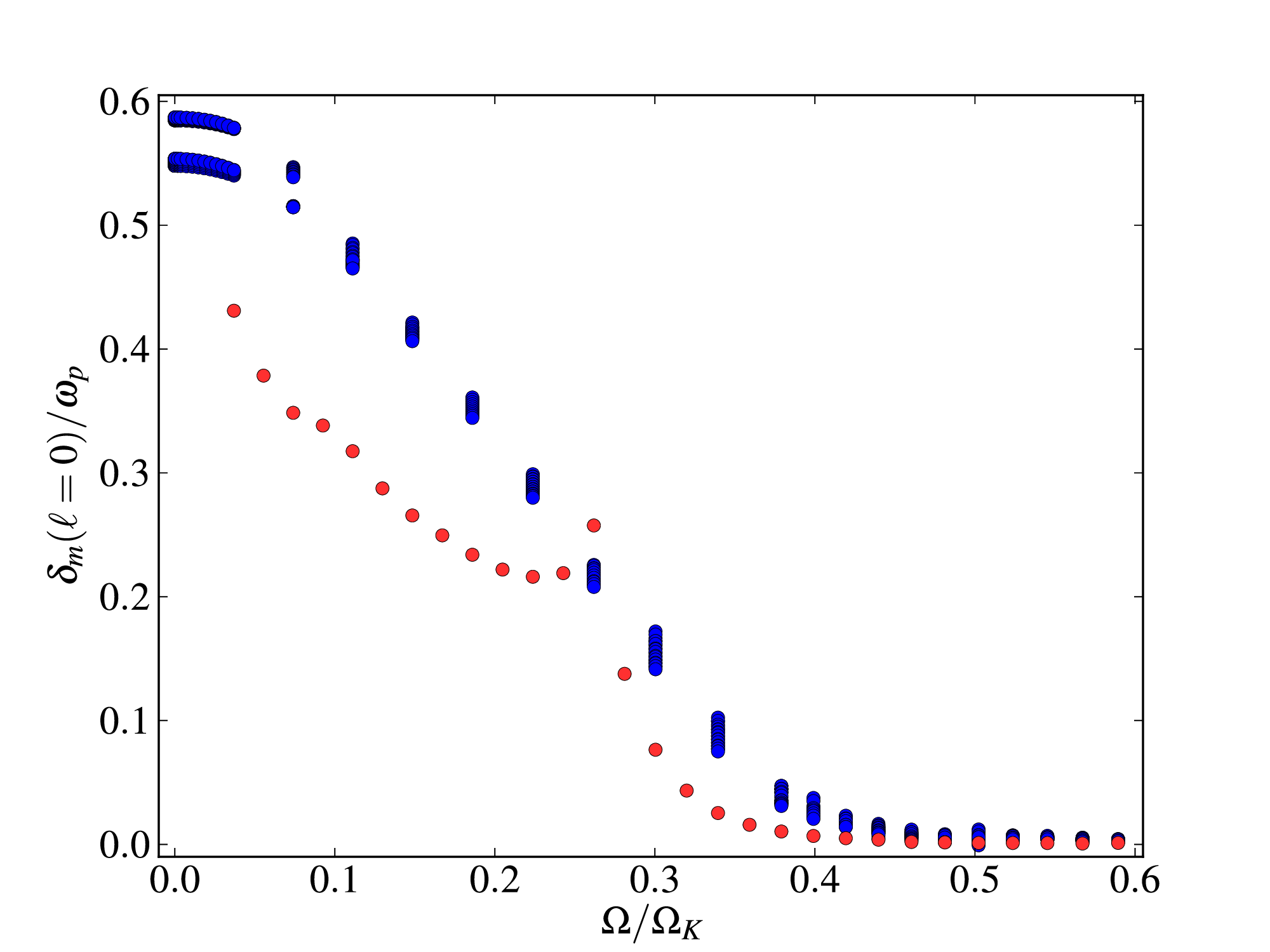}
   \caption{(Colour online) Comparison of frequency spacing $\delta_m$ between island modes, computed from numerical simulations and semi-analytical formula Eq. (\ref{omen0}) for different values of $\Omega/\Omega_K$ (the frequency spacings are normalized by $\omega_p=\sqrt{GM/R_p^3}$ with $R_p$ the polar radius). Red/dark gray: semi-analytical results of Eq. (\ref{omen0}) for $n=42$ and $m=1$, blue/black: numerical results. The numerical results correspond to different sets of $\delta_m$ for $n\in [42,51]$, $\ell=0$ and $m\in\{-1,1\}$.}
   \label{regfig2}
\end{figure}

\begin{figure}
  \centering
    \includegraphics[width=\linewidth]{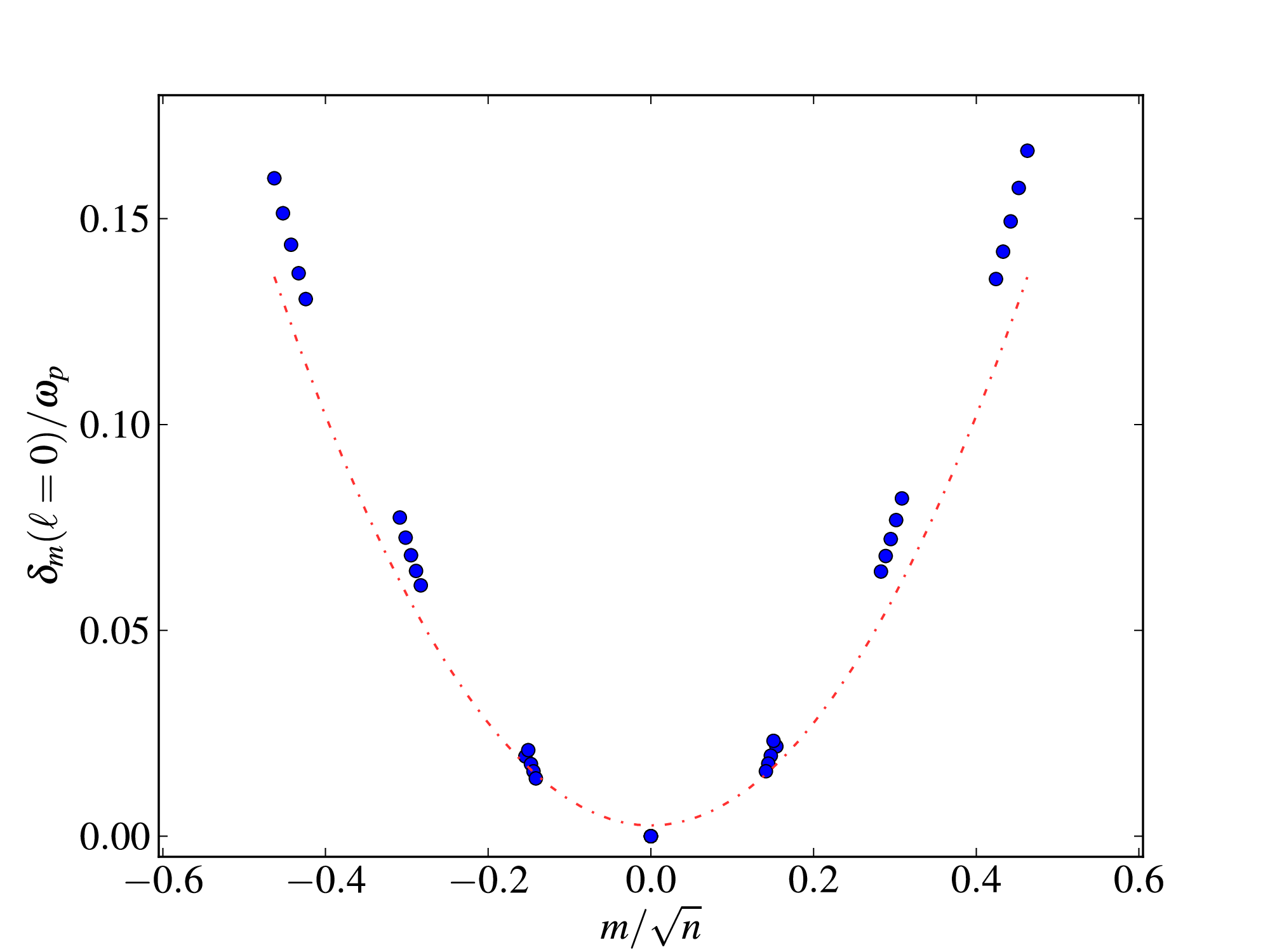}
\caption{(Colour online) Frequency spacings $\delta_m (\ell=0) =\omega_{n,\ell=0,m}-\omega_{n,\ell=0,m=0}$, normalized by $\omega_p$, as a function of $m/\sqrt{n}$. The integers $n$, $\ell$ and $m$ are the quantum numbers of island modes. The rotation rate is $\Omega/\Omega_K=0.419$. Blue/black dots: numerical results. Red/dark gray dashed line: semi-analytical results for the last term in Eq. (\ref{omen3}). Numerical modes included are for quantum numbers $n\in \{42, 44, 46, 48, 50\}$, $\ell=0$ and $m\in[-3,3]$.}
\label{regum}
 \end{figure}

In order to investigate the drift between the spectra of different $m$ as rotation increases, we consider the frequency spacing:
\begin{equation}
 \delta_m = \omega_{n, \ell, m}-\omega_{n, \ell, 0}~.
\end{equation}
Figure \ref{regfig2} displays a comparison between the numerical and semi-analytical values of $\delta_m$ for $\ell=0$ and $|m|=1$. 
As expected, the agreement is not good at small rotation rates.
Using the first order of Tassoul's formula to approximate the numerical results at zero rotation, 
$\delta_m^N=\omega^N_{n, \ell, m}-\omega^N_{n, \ell, 0}$ is found to be close to $|m| \Delta/2$ when $\Omega/\Omega_K=0$. 
This is not compatible with the asymptotic theory of the island mode since it predicts that $\delta_m$ goes to zero when $\omega$ goes to infinity.
Indeed, $\delta_n(m)$ depends on $m/\omega$ because $\tilde{c}_s$ and the ray path $\gamma$, given by the Hamiltonian Eq. (\ref{wkb}), both depend on $m/\omega$. 
An alternative explanation is to consider the spatial distribution of island modes of fixed $m$ and $\ell$: one finds that increasing $\omega$ produces both larger derivatives along the stable ray associated
with a higher node number $n$, and larger transverse derivatives because the transverse extent scales as $1/\sqrt{\omega}$. 
Thus, the contribution of the azimuthal derivatives becomes negligible in the wave equation Eq. (\ref{helm}). 
We also verified that $\delta_m$ displayed in Fig. \ref{regfig2} diminishes when $n$ is increased.
Thus, for rotation rates such that the numerical modes are not fully island modes, they behave more like spherical modes
and $\delta_m^N$ shows clear discrepancies with the asymptotic results.
 By contrast, at high rotation rates, an approximate analytical formula for $\delta_m$ is derived in the following and shown to closely reproduce the numerical results.
Starting from 
\begin{equation}
\label{omen0}
 \delta_m (\ell=0) = [\delta_n(m)-\delta_n(0)] n + [\beta(m)-\beta(0)]~,
\end{equation}
we first assume that $n$ is large enough to neglect $\beta(m)-\beta(0)$. 
From Eq. (\ref{reg2}), $\delta_n(m)$ is equal to $2 \pi/T_\gamma(m)$ where
\begin{equation}
\label{timeint}
 T_\gamma(m)=\oint_\gamma \frac{ds}{c_s}\sqrt{1-\frac{1}{\omega^2} \left[ \omega_c^2+ \frac{c_s^2\left( m^2 - \frac{1}{4}\right)}{d^2}\right]}~.
\end{equation}
The dependence of $T_\gamma$ in $m$ is explicit in the integrand but is implicit in the integration path $\gamma$. 
In the following, the variation of the location of $\gamma$ with $m/\omega$ is assumed to be negligible.
Then, an expansion in $1/\omega$ of the integrand in Eq. (\ref{timeint}) leads to:
\begin{equation}
 T_\gamma(m) \simeq \oint_\gamma\frac{ds}{c_s} - \frac{1}{2\omega^2}\left[\oint_\gamma\omega_c^2\frac{ds}{c_s} + \oint_\gamma \frac{c_s\left(m^2-1/4\right)}{d^2}ds\right]~.
\end{equation}
Hence we obtain an approximate expression for $\delta_n(m)$ of the form:
\begin{equation}
\label{delnoui}
\delta_n(m) \simeq \frac{2\pi}{\oint_\gamma\frac{ds}{c_s}} + \frac{\pi}{\omega^2} \frac{\oint_\gamma \frac{\omega_c^2}{c_s} ds + \oint_\gamma \frac{c_s\left(m^2-1/4\right)}{d^2}ds} {\left(\oint_\gamma\frac{ds}{c_s}\right)^2}~.
\end{equation}
If we insert the previous expression for $\delta_n(m)$ in Eq. (\ref{omen0}) and neglect $\beta(m)-\beta(0)$, we have
\begin{equation}
\label{omen1}
 \delta_m (\ell=0) \simeq \left[ \frac{m^2}{\omega^2} \pi \frac{\oint_\gamma \frac{c_s}{d^2}ds}{\left(\oint_\gamma\frac{ds}{c_s}\right)^2} \right] n~.
\end{equation}
Finally, normalizing by $\omega_p=\sqrt{GM/R_p^3}$ and replacing $\omega$ by $n \frac{\omega}{n}$ yields
\begin{equation}
\label{omen3}
\begin{split}
 \frac{\delta_m (\ell=0)}{\omega_p}& \simeq \left( \frac{m}{\sqrt{n}}\right)^2 \pi \frac{\oint_\gamma \frac{c_s}{d^2}ds}{\left(\oint_\gamma\frac{ds}{c_s}\right)^2} \left(\frac{n}{\omega/\omega_p}\right)^2 \frac{1}{\omega_p^3}\\
 & \simeq \left( \frac{m}{\sqrt{n}}\right)^2 \frac{1}{4\pi\omega_p} \oint_\gamma \frac{c_s}{d^2}ds~,
 \end{split}
\end{equation}
where we have used the fact that $\frac{n}{\omega/\omega_p}$ stays constant in the frequency range considered here, and is known to be close to $\omega_p \oint_\gamma\frac{ds}{c_s}/2\pi$. 
The previous expression will be made more precise by renormalizing the value of $c_s$ by $\sqrt{1-\omega_c^2/\omega^2}$ to take into account that $\omega_c$ is not negligible, and indeed is of the order of $\omega$, close to the stellar surface. 
In Fig. \ref{regum}, the numerical values of $\delta_m(\ell=0)$ as well as results for the last term in Eq. (\ref{omen3}) are plotted as a function of $m/\sqrt{n}$ for $\Omega/\Omega_K = 0.419$, showing a good agreement.
This behavior is valid for rotation rates higher than $\Omega/\Omega_K \simeq 0.4$.
It must also be noted that in the numerical calculations by \citet{reese2009}, using more realistic stellar models, the asymptotic
$\frac{m}{\sqrt{n}}$ dependency was also found empirically.


\subsection{Pressure amplitudes of island modes}
\label{compnumamp}
\begin{figure*}
   \centering
    \includegraphics[width=\linewidth]{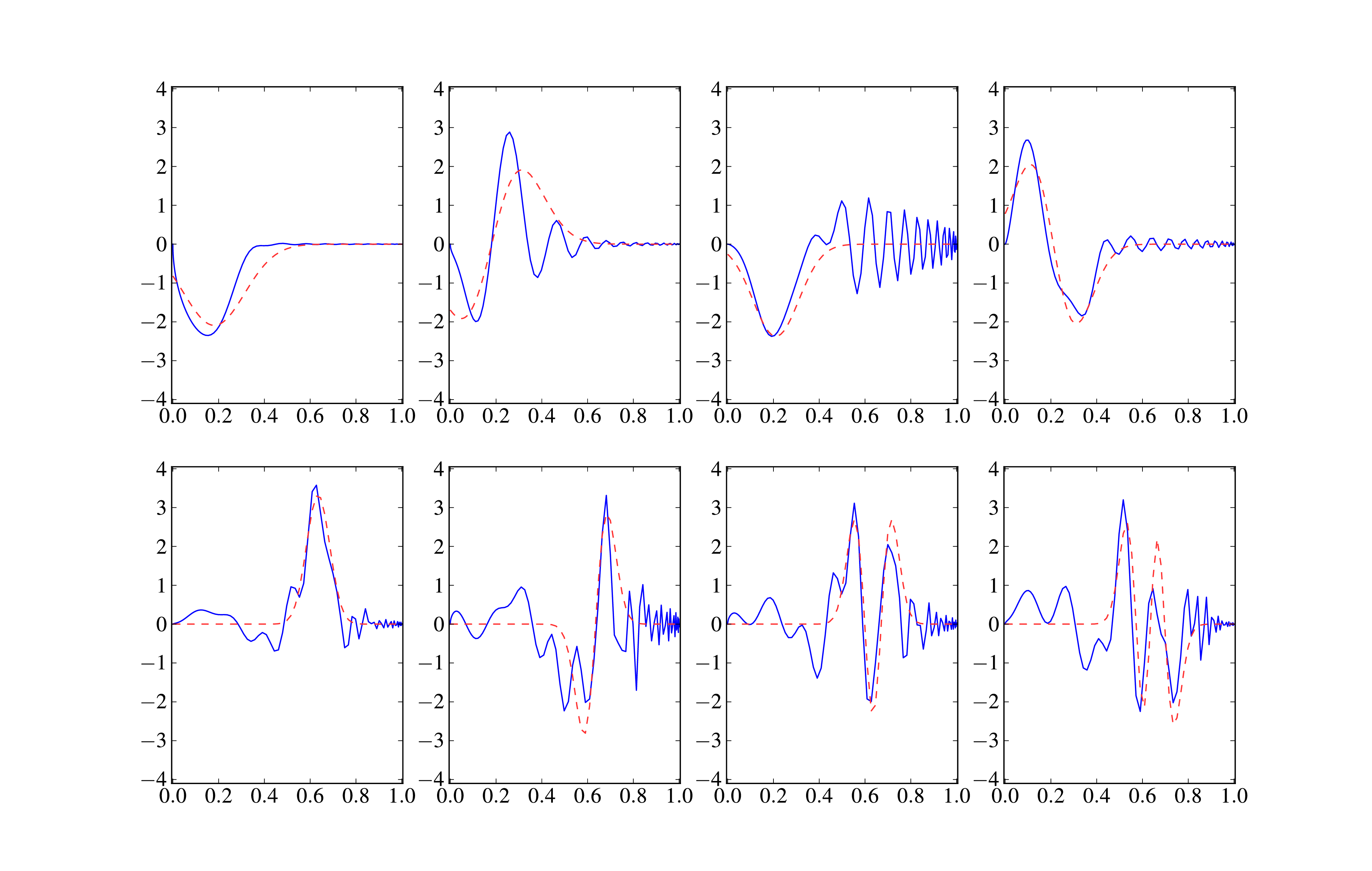}    
   \caption{(Colour online) Examples of normalized amplitudes distributions (real part  of $\Phi_m^\ell$) on the equator as a function of position $r/R_{eq}$ (where $R_{eq}$ is the equatorial radius of the stellar model) for semi-analytical and numerical modes. Modes displayed are for $\Omega/\Omega_K=0.300$ (upper line), and $\Omega/\Omega_K=0.707$ (lower line). Quantum numbers $(n, \ell, m)$ are as follows: upper line $(50,0,0)$, $(50,1,0)$, $(50,0,1)$, $(50,1,1)$. 
Lower line $(50,0,0)$, $(50,1,0)$, $(50,2,0)$, $(50,3,0)$. Blue/black continuous line: numerical results; red/dark gray dashed line: semi-analytical results.}
   \label{modeq}
\end{figure*}

In this section, we compare the results obtained from the semi-analytical formula for mode spatial distributions Eq. (\ref{modefull}) with results from full numerical computations. 
Equatorial cuts of the semi-analytical modes can be expressed as a function of $\nu=\sqrt{\omega} (r-r_0)$, where $r_0$ is the radial position of the ray $\gamma$,
while the value of $\Gamma$ is obtained from the eigenvectors of the monodromy matrix $M$.
In Fig. \ref{modeq}, the equatorial cuts of semi-analytical and numerical modes are plotted for different rotational velocities and quantum numbers $\ell$ and $m$. 
The chosen modes are representative of the different behaviors observed.
Discrepancies between semi-analytical and numerical results are mainly due to edge effects. 
This occurs when the transverse extent of the mode (which scales as $1/\sqrt{\omega}$) reaches either the polar axis for small rotations, or the surface near the equator for high rotations.
Finally, avoided crossings can also contravene an accurate prediction for mode amplitudes since the amplitudes of crossing modes will be linear combinations of all the modes contributing to the crossing. Hence, modes undergoing an avoided crossing can differ significantly from Eq. (\ref{modefull}) (cf. third panel in Fig. \ref{modeq}). Overall, there is nevertheless a good agreement between the semi-analytical and numerical results for mode spatial distributions, showing the validity of Eq. (\ref{modefull}).


\section{Phenomenology and observables for asteroseismology}
\label{pheno}
In this section, we show that the asymptotic theory provides a simple understanding of the evolution of the island mode spectrum with rotation.
Then, the physical content of the potentially observable frequency spacings $\delta_n$, $\delta_\ell$ and $\delta_m$ is discussed.

\begin{figure}
   \centering
   \includegraphics[width=\linewidth]{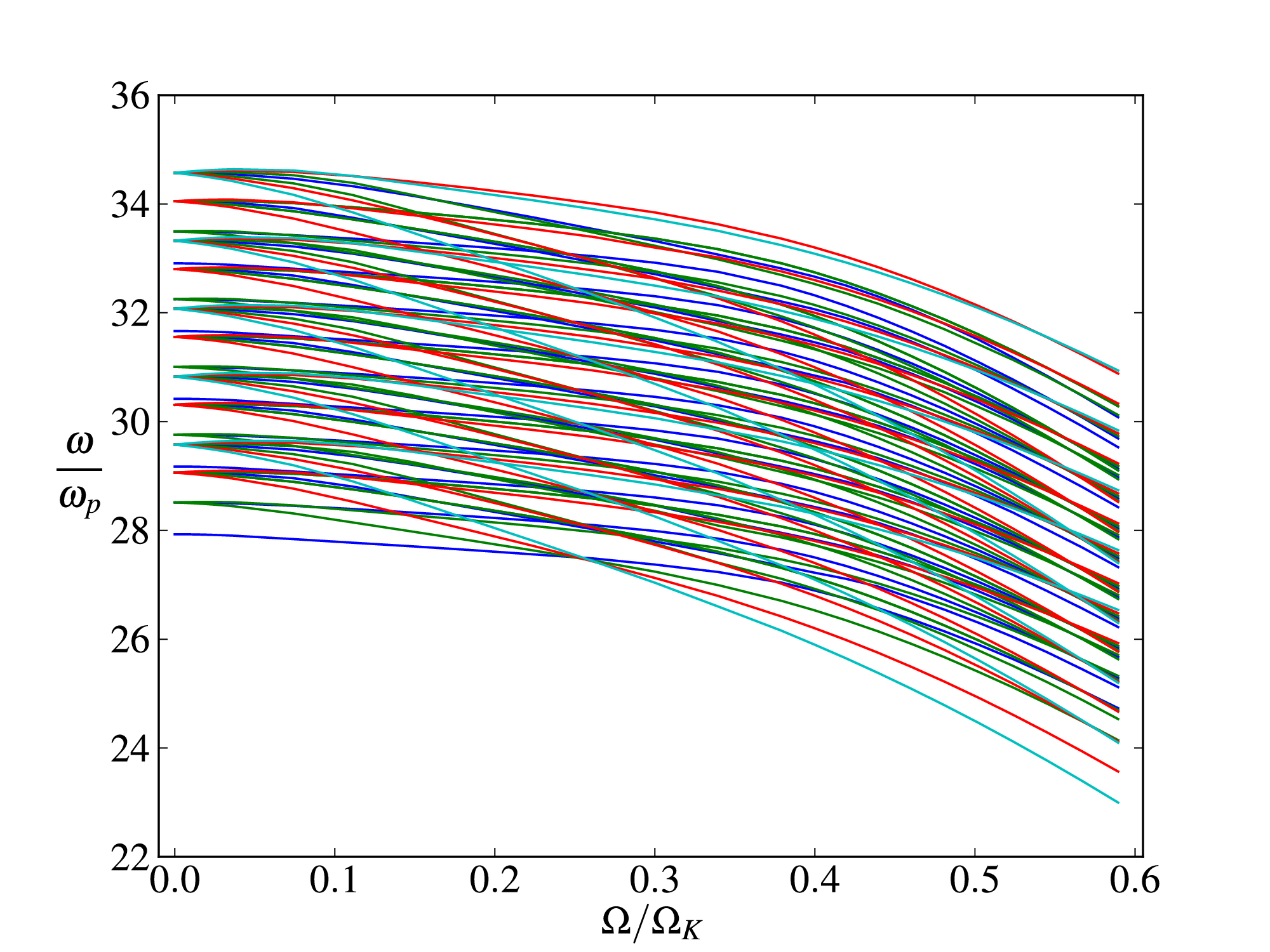}
   \caption{(Colour online) Normalized frequencies $\omega/\omega_p$ as a function of normalized rotation velocity $\Omega/\Omega_K$ in the observer's frame. Different colors correspond to different values of $|m|$. Blue: $m=0$, green: $|m|=1$, red: $|m|=2$, cyan: $|m|=3$. Modes included are for the quantum numbers listed in Table \ref{qunumtab} (or equivalently $n_s \in[21,25]$, $\ell_s \in[0,3]$, $m_s \in[-\ell_s,\ell_s]$)}
   \label{freqs}
\end{figure}

\begin{figure}
   \centering
   \includegraphics[width=\linewidth]{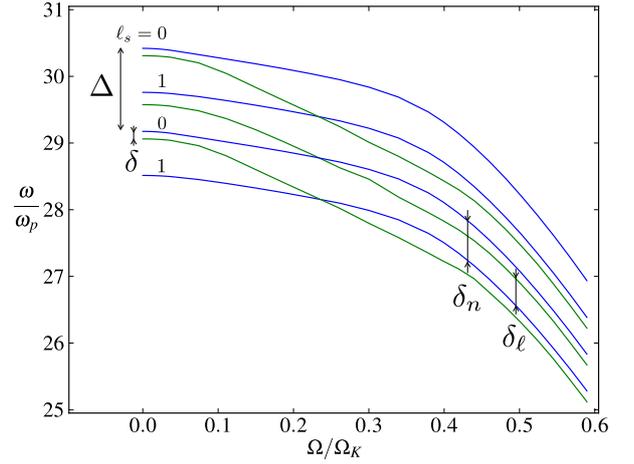}
   \caption{(Colour online) Normalized frequencies $\omega/\omega_p$ as a function of normalized rotation velocity $\Omega/\Omega_K$.
   The quantum numbers of the displayed modes are: $\ell=0$, $n\in [43,46]$ and $\ell=1$, $n\in [42,44]$ with $m=0$.
  Blue: $\ell=0$, green: $\ell=1$.  
 For clarity, the corresponding degrees of the spherical harmonics are also written.
 We have outlined the large frequency separation $\Delta$, the small frequency separation $\delta = \omega_{n_s,\ell_s} - \omega_{n_s-1,\ell_s+2}$ and the $\delta_n$, $\delta_\ell$ spacings with arrows.}
   \label{smallsep}
\end{figure}

Figure \ref{freqs} displays the global evolution of all the numerical frequencies considered in the observer's frame, whose island mode quantum numbers can be found in Table \ref{qunumtab} 
(or equivalently $n_s \in[21,25]$, $\ell_s \in[0,3]$, $m_s \in[-\ell_s,\ell_s]$ in spherical modes quantum numbers).
The first phenomenon that can be noticed is a global decrease of frequencies with rotation. This effect is simply due to the increasing volume of the star when it is spinning rapidly.
Besides this global effect, the evolution of the spectrum's organization can be inferred from the evolution of frequency spacings $\delta_n$, $\delta_\ell$ and $\delta_m$. 
The spacing $\delta_n$ stays almost constant from null up to high rotations, its value remaining close to half the large 
frequency separation of the spherical model. 
If a large number of island modes are detected in an observed spectrum, $\delta_n$ should be easily extracted from the data. 
By contrast, the rapid evolution of $\delta_\ell$ with rotation will strongly modify the spectrum's organization. 
This is shown in Fig. \ref{smallsep} where for clarity only a few $m=0$ modes have been displayed: the $\ell=0,n\in [43,46]$ and $\ell=1,n\in [42,44]$
modes (or equivalently the $(n_s, \ell_s) \in \{(21,1), (21,2), (21,3), (22,0), (22,1), (22,2), (23,0)\}$ modes).
Starting from the usual structure at zero rotation involving the large and small separations of Tassoul's theory, the spectrum reorganization 
induced by the decrease of $\delta_\ell$ can also be viewed as an increase of the small separation $\delta$.
Then, above $\Omega/\Omega_K \simeq 0.45$, the structure of the $m=0$ spectrum remains practically unchanged.

\begin{figure}
   \centering
   \includegraphics[width=\linewidth]{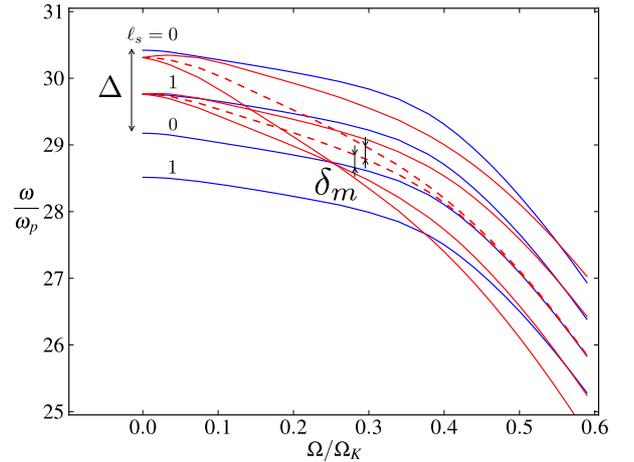}
   \caption{(Colour online) Normalized frequencies $\omega/\omega_p$ as a function of normalized rotation velocity $\Omega/\Omega_K$. 
  The quantum numbers of the displayed modes are: $n=44$, $\ell=0$, $m \in \{-2,-1,1,2\}$ with the $n\in [43,46]$, $\ell=0$, $m=0$ modes.
  Blue: $m=0$, red: $|m|\in\{1,2\}$. 
 For clarity, the corresponding degrees of the spherical harmonics are also written.
 Continuous line: frequencies in the observer frame, dashed line: in the inertial frame.
 We have outlined the large frequency separation $\Delta$ and the $\delta_m$ spacing with arrows.}
   \label{rotsplit}
\end{figure}

Now, to illustrate the evolution of the spectra of different $m$, 
Fig. \ref{rotsplit} displays the $n=44$, $\ell=0$, $m \in \{-2,-1,1,2\}$ mode frequencies as a function of the rotation rate 
together with the $n\in [43,46]$, $\ell=0$, $m=0$ frequencies.
The main feature of this evolution is the decrease of $\delta_m$ from $\simeq \frac{\Delta}{2}|m|$ at zero rotation to very small
values at high rotations. 
When multiplets of island modes are defined as in Sect. \ref{compnum} and Fig. \ref{node}, they show no regularity at small rotation rates. 
Note, however, that the splitting $\omega_{m}-\omega_{-m}$ is always very close to $-2m\Omega$ because the effects of the Coriolis force are negligible.
By contrast, at high rotation rates, as $\delta_m$ vanishes above $\Omega/\Omega_K \simeq 0.45$, the  $m \in \{-2,-1,0,1,2\}$ modes clearly form a regular multiplet, as can be seen in Fig. \ref{rotsplit},
where deviations from strict $\Omega$ spacings are due to the $(m/\sqrt{n})^2$ term. 
Since for such rotation rates the structure of the $m=0$ spectrum remains unchanged, the evolution of the whole spectrum in the observer's frame is dominated by the advection term $m\Omega$.

The global evolution of mode frequencies in the observer's frame shown in Fig. \ref{freqs} also presents some particular events: a first clustering of mode frequencies occurs around $\Omega/\Omega_K\simeq 0.25$ and then a second one around $\Omega/\Omega_K\simeq 0.56$.
Both phenomena can be understood from the asymptotic theory.
According to the asymptotic formulas  Eqs. (\ref{regul}-\ref{reg3}), crossings of mode frequencies will happen when $\delta_\ell/\delta_n$, or equivalently $\alpha/\pi$, has a rational value.
Though the asymptotic theory predicts true eigenvalue crossings, it is known that these crossings will be avoided if the two modes are of the same symmetry class \citep{lali}.
As can be seen on Fig. \ref{regfig}, $\delta_\ell$ becomes equal to $\delta_n$ at some rotation rate around $\Omega/\Omega_K\simeq 0.25$ where the spectrum for a given $m$
simplifies to $\omega_{n,m} = \delta_n(m) n + \beta$. 
The degeneracy occurs between modes of a different symmetry class, and the rotation rate at which it occurs depends only weakly on the $m$ values of the modes, if $m$ is small. 
This property translates itself into a clustering of the full spectrum in the observer's frame because
it turns out that this rotation rate is close to $\delta_n/2$; and $\delta_m$, that decreases from an initial value of $m \delta_n$ to zero at high rotation, is around $m \delta_n/2$ at this intermediate rotation. 
The second frequency clustering close to $\Omega/\Omega_K\simeq 0.56$ is related to the fact that $\delta_m$ vanishes at high rotation. In this regime,
the different $m$ spectra are expected to collapse onto a single spectrum in the rotating frame but not in the observer's frame. However,
when $\Omega$ is equal to $\delta_n$, the near degeneracy of the $m$ spectra produces the frequency clustering observed at $\Omega/\Omega_K\simeq 0.56$.

One of the interests of asymptotic theories in asteroseismology is to gain physical insights into seismic observables 
such as $\delta_n$, $\delta_l$, and $\delta_m$. In the following, we briefly discuss this point with emphasis on the differences and similarities with
the physical content of the large and small separation from Tassoul's theory.
The spacing $\delta_n$ depends only on the acoustic travel time $T_\gamma= \oint_\gamma \frac{ds}{\tilde{c}_s}$ along the acoustic ray $\gamma$. 
We expect $T_\gamma$ to be dominated by the time spent in the sub-surface region where the sound speed is much smaller
than in the interior. While the path of the ray varies with rotation, $\delta_n$ remains approximately proportional to the mean density as 
shown by \citet{retal2008}.
On the other hand, the $\delta_\ell$ spacing depends also on the second derivatives of the sound speed transverse to the ray $\gamma$, an information integrated all along the ray. 
As long as the path of the stable ray goes through the central region of the star, the island mode frequencies should be sensitive 
to the chemical stratification and thus the age of the star.
However, after the bifurcation at $\Omega/\Omega_K\simeq 0.26$, the ray path progressively avoids the central region and the island modes do not contain this
information anymore.
Another interesting property of $\delta_\ell$ (or $\delta_\ell / \delta_n$) is that it is very sensitive to rotation as long as $\Omega/\Omega_K \leq 0.35$. 
Finally, for high rotation rates ($\Omega/\Omega_K\geq 0.40$), the value of $\delta_m$, that can be detected through the irregularity of multiplets, also gives an information on rotation 
since it is proportional to $\oint_\gamma \frac{c_s}{d^2}ds$ (Eq. (\ref{omen3})) where the distance of the ray to the rotation axis $d$ strongly depends on the rotation rate.


\section{Conclusions}
In this paper, we derived an asymptotic formula for frequencies that predicts and describes regular spacings in the p-mode spectrum of rapidly rotating stars.
The derivation relied on finding a stable periodic solution of the acoustic ray dynamics, and obtaining an expression for the modes that are localized around this ray, the so-called island modes.
The method thus provides a formula for the island modes frequencies, as well as a formula for the mode spatial distributions.
We compared these semi-analytical formulas with results from numerical computations of high-frequency oscillations in rotating polytropic stellar models. 
The frequency spectrum is characterized by the three spacings $\delta_n$, $\delta_\ell$ and $\delta_m$.
The agreement was shown to be good for $\delta_n$ and $\delta_\ell$ at almost all rotation rates, while $\delta_m$ shows significant discrepancies at low rotation rates.
The spacing $\delta_n$ stays almost constant at all rotation rates with a value that is close to half the large frequency separation of the non-rotating model. 
On the other hand, the rapid decrease of $\delta_\ell$ strongly modifies the spectrum's organization up to $\Omega/\Omega_K\simeq0.4$, while above that rotation rate, $\delta_\ell$ remains approximately constant.
For such high rotation rates, the spacing $\delta_m$ nearly vanishes, thus
in the observer's frame the evolution of the whole spectrum is dominated by the advection term $m \Omega$.
We have also seen that the combined evolution of these frequency spacings with rotation leads to particular events such as true or near degeneracies, that can significantly simplify the spectrum.
In addition to these new insights on the evolution of the island mode spectrum with rotation, the asymptotic theory provides semi-analytical formulas for the regular spacings, in particular simple formulas for $\delta_n$ and $\delta_\ell$.

The present asymptotic theory should be useful for different aspects of stellar seismology in the presence of rapid stellar rotation.
The regular frequency spacings are potentially observable and our results provide guidance to look for them in data. 
While investigations dedicated to the search for regularities are necessary (e.g. \citet{letal2010}), we expect that
the easiest quantities to detect in an island mode spectrum are $\delta_n$ at any rotation rate, $2m\Omega$ at small rotation rates, and $\Omega$ at high rotation.
For modeling pulsations, the asymptotic theory provides a new approach, complementary to numerical computations. One of its advantages
is to give a quick estimate of frequency spacings for a given stellar model, which in turn can be used to search for patterns in numerically computed spectra.
In the same spirit, the semi-analytical amplitude distributions might provide a useful approximation for calculating mode visibilities and spectral signatures.

The asymptotic theory in itself can be improved and extended in various ways.
We have already mentioned that the method needs to be refined at the rotation rate where the bifurcation of the stable ray occurs, using
uniform approximations of the ray dynamics. It would also be interesting to predict analytically the rotation rate of the bifurcation for a given sequence of stellar models.
The present method also assumes that the modes are governed by local dynamics around the stable ray. This assumption can be tested with a numerical EBK method applied to the tori
of the stable island \citep{bohigas93}, although this method
is complicated to implement in pratice.
In this paper, we left aside the determination of the actual number of modes that are described by the asymptotic theory. An estimate of such a number can be obtained by computing systematically the phase space volume of stable islands for different rotations (e.g. \citet{lg2009}).
Then, knowing the value of $\delta_\ell$ that gives the mean distance between island mode frequencies, or the mean density of these modes, one could compute the $\ell^{\mathrm{max}}$ of the modes that satisfy our formulas.
Another aspect that we have not modeled is avoided crossing in spite of the fact that it will induce important deviations, especially at low frequencies. 
Strong gradients of the sound speed will also produce deviations from the asymptotic theory. A technique called ray-splitting, that has already been used successfully in quantum chaos \citep{blumel96},
could account for this effect.
Finally, a similar technique could be applied to asymptotic gravity modes, that were shown recently to have connections with ray theory \citep{ballot2011}.


\begin{acknowledgements}
We thank J. Ballot for his help at various stages of this work. 
We also thank the ANR project SIROCO for funding and CALMIP (``CALcul en MIdi-Pyr\'en\'ees'') for the use of their supercomputer. 
M.P., F.L. and D.R.R. acknowledge the KITP staff of UCSB for their warm hospitality during the research program ``Asteroseismology in the Space Age''. 
This research was supported in part by the National Science Foundation under Grant No. NSF PHY05--51164.
D.R.R. acknowledges financial support through a postdoctoral fellowship from the ``Subside f\'ed\'eral pour la recherche 2011'', University of Li\`ege.
\end{acknowledgements}


\begin{appendix} 
\section{Wave equation in the meridional plane}
\label{2D}
In this section, we derive the two-dimensional wave equation in the meridional plane of the star. 
We start from the three-dimensional wave equation Eq. (\ref{helm}) in spherical coordinates $(r,\theta,\phi)$, and from the expression of the mode amplitude as $\Psi = \Psi_m \exp ( i m \phi)$ we obtain the following wave equation:
\begin{equation}
\label{appen1}
 \left[ \Delta_{r,\theta} + \frac{1}{r}\left(\frac{\partial }{\partial r} + \frac{1}{r \tan \theta} \frac{\partial }{\partial \theta} \right) - \frac{m^2}{(r\sin\theta)^2} \right]\Psi_m + \frac{\omega ^2 - \omega_c ^2}{c_s ^2} \Psi_m = 0~,
\end{equation}
where $\theta$ is the colatitude. 
We want to cancel out of this equation the terms multiplied by first order derivatives of $\Psi_m$ in order to obtain a two-dimensional Helmholtz-like equation. 
We thus introduce the ansatz $\Psi_m(r,\theta)=\beta(r,\theta) \Phi_m (r,\theta)$ in Eq. (\ref{appen1}) to obtain that $\beta(r,\theta)$ must satisfy
   \begin{equation}
      \frac{2}{\beta}\frac{\partial \beta}{\partial r}  +  \frac{1}{r}  =  0 
 \end{equation}
to cancel out terms in $ \frac{\partial \Phi_m}{\partial r} $, and 
\begin{equation}
      \frac{2}{\beta}\frac{\partial \beta}{\partial \theta}  +  \frac{1}{\tan \theta}   =  0
   \end{equation}
to cancel out terms in $ \frac{\partial \Phi_m}{\partial \theta} $. 
This leads to the solution
\begin{equation}
 \beta(r,\theta) =  \frac{B}{\sqrt{r \sin \theta}},
\end{equation}
and we choose $B=1$ to yield Eq. (\ref{helm2}). 

\section{Expression of the imaginary part of $\Gamma$}
\label{wro}
In this section, we derive the expression for the imaginary part of $\Gamma$ using the Wronskian of Eqs. (\ref{ham1}-\ref{ham2}).
For these equations, the Wronskian is defined as
\begin{equation}
\label{wrons}
 W(s) = z(s) \frac{1}{\tilde{c}_s(s)}\frac{d \bar{z}(s)}{d s} - \frac{1}{\tilde{c}_s(s)}\frac{d z(s)}{d s}\bar{z}(s) ~,
\end{equation}
corresponding to
\begin{equation}
 W(s) = z(s) \bar{p}(s) - p(s) \bar{z}(s)~,
 \end{equation}
where $(z,p)$ and $(\bar{z},\bar{p})$ are the two independent solutions of Eqs. (\ref{ham1}-\ref{ham2}).
Now, in order to obtain the variation of the Wronskian with $s$, we  need the Abel's differential equation identity that says that for an equation of the following type
\begin{equation}
\label{appen2}
 \frac{d^2 y}{d x ^2}+ P(x) \frac{dy}{dx} + Q(x) y = 0~,
\end{equation}
 with the Wronskian 
\begin{equation}
 W=y_1 \frac{d y_2}{dx} - \frac{dy_1}{dx} y_2~,
\end{equation}
and $y_1$, $y_2$ the two independent solutions of Eq. (\ref{appen2}), there is the identity
\begin{equation}
 W(x) = W_0 \exp \left(- \int^x P(x') dx' \right)~.
\end{equation}
If we apply this identity to the equation on $z(s)$ that can be derived from Eqs. (\ref{ham1}-\ref{ham2}) we obtain that
\begin{equation}
 W(s) = W(s_0)\exp \left( - \int_{s_0}^{s} 0 \times \tilde{c}_s~ ds'\right)~.
\end{equation}
This yields that the Wronskian stays constant for all $s$:
\begin{equation}
 W(s) = W(s_0)=W_0~. 
 \label{beta}
 \end{equation}
From Eqs. (\ref{ricca}), (\ref{wrons}) and (\ref{beta}) we thus obtain that
\begin{equation}
 \bar{\Gamma} - \Gamma = \frac{W_0}{z \bar{z}} ~,
\end{equation}
yielding
\begin{equation}
 \mathrm{Im}[\Gamma] = \frac{i}{2} \left[\frac{W_0}{z \bar{z}}\right]~,
\end{equation}
and therefore
\begin{equation}
 \mathrm{Im}[\Gamma] = \frac{i}{2} \left[\frac{W_0}{|z|^2}\right] ~.
\end{equation}
Hence this shows that $\mathrm{Im}[\Gamma]$ keeps a constant sign along the ray when $s$ varies. 
Thus if $\Gamma$ is chosen such that it localizes the function at $s_0$, then the function will stay localized around the ray for all $s$. 
Since the localization is obtained for $\mathrm{Im}[\Gamma] > 0$ or equivalently $i W_0 > 0$, we choose the Wronskian such that
\begin{equation}
 W(s) = - i~.
\end{equation}


\section{Analogy between the function $z$ and the deviation of nearby rays}
\label{jaco}
In this section, we show that the equation satisfied by the function $z$ corresponding to Eqs. (\ref{ham1}-\ref{ham2}) is the same as the equation describing the deviation of two nearby rays.
It will thus be possible to compute the evolution of $z$ from the ray dynamics.
To be an acoustic ray of the system, the ray $\gamma$ must correspond to an extremum of the action $S$ defined as
\begin{equation}
 S(q_i, q_f, \omega)=\int^{q_f}_{q_i} \vec{p} \cdot d\vec{q} = \int^{q_f}_{q_i} \frac{1}{\tilde{c}_s} d \sigma~.
\end{equation}
The length element $d\sigma$ can be expressed in the coordinate system centered on the ray $\gamma$ defined in Sect. \ref{harmeq} as
\begin{equation}
 d \sigma ^2= ( 1 - \xi \kappa (s) ) ^2 d s  ^2 + d\xi ^2~,
\end{equation}
yielding
\begin{equation}
 d \sigma ^2= \left[ ( 1 - \xi \kappa (s) ) ^2 + \left( \frac{d\xi}{ds} \right) ^2 \right] d s ^2~.
\end{equation}
The action function $S$ can therefore be expressed in this coordinate system as
\begin{equation}
 S =\int \frac{1}{\tilde{c}_s(s,\xi)} \sqrt{\left[ ( 1 - \xi \kappa (s) ) ^2 + \left(\frac{d\xi}{ds}\right) ^2 \right]} d s~,
\end{equation}
with the corresponding Lagrangian
\begin{equation}
 L (s, \xi ,d\xi/ds) = \frac{1}{\tilde{c}_s(s,\xi)} \sqrt{\left[ ( 1 - \xi \kappa (s) ) ^2 + \left( \frac{d\xi}{ds} \right) ^2 \right]}
\end{equation}
The Lagrangian $L$ becomes for $\xi$ small:
\begin{equation}
 \begin{split}
 L (s, \xi ,\dot{\xi}) = \frac{1}{\tilde{c}_s(s)} \left[ 1- \xi \kappa(s) + \frac{1}{2} \dot{\xi} ^2  - \frac{1}{\tilde{c}_s(s)} \frac{\partial \tilde{c}_s}{\partial \xi}  \xi\right. \\
\left. + \frac{\kappa(s)}{\tilde{c}_s(s)} \frac{\partial \tilde{c}_s}{\partial \xi}  \xi^2  - \frac{1}{2 \tilde{c}_s (s)} \frac{\partial ^2 \tilde {c}_s}{\partial \xi ^2} \xi ^2 \right. \\
\left. + \frac{1}{\tilde{c}_s(s) ^2} \left( \frac{\partial \tilde{c}_s}{\partial \xi}\right)^2 \xi ^2 \right] + O(\xi^3)~,
\end{split}
\end{equation}
where $\dot{\xi} \equiv d\xi/ds$ and derivatives in $\xi$ are evaluated at $\xi=0$.
As is known from classical mechanics, for the action to be extremum, the associated Lagrangian must satisfy the Euler-Lagrange equation
\begin{equation}
 \frac{d}{d s} \left( \frac{\partial L}{\partial \dot{\xi} }\right) - \frac{\partial L}{\partial \xi} = 0~.
\end{equation}
Keeping only quadratic terms in $\xi$ from this equation yields
\begin{equation}
\begin{split}
 \ddot{\xi} - \frac{1}{\tilde{c}_s(s)}\frac{d \tilde{c}_s(s)}{d s} \dot{\xi} - 2 \left[ \frac{1}{\tilde{c}_s(s) ^2} \left(\frac{\partial \tilde{c}_s}{\partial \xi} \right) ^2 + \frac{\kappa (s)}{\tilde{c}_s(s)}\frac{\partial \tilde{c}_s}{\partial \xi} \right. \\
\left.- \frac{1}{2 \tilde{c}_s(s)} \frac{ \partial ^2 \tilde{c}_s}{\partial \xi ^2} \right] \xi + \left[\kappa (s) + \frac{1}{\tilde{c}_s(s)} \frac{\partial \tilde{c}_s}{\partial \xi} \right] + O(\xi^3) = 0~.
\end{split}
\end{equation}
This equality must be valid for all $\xi$ so we can deduce that:
\begin{equation}
 \kappa (s) = - \frac{1}{\tilde{c}_s(s)} \frac{\partial \tilde{c}_s}{\partial \xi}~,
\end{equation}
and
\begin{equation}
\begin{split}
 \frac{1}{\tilde{c}_s(s) } \frac{d }{d s } \left( \frac{1}{\tilde{c}_s(s) }\frac{d \xi}{d s} \right) - \frac{2}{\tilde{c}_s(s) ^2} \left[ \frac{1}{\tilde{c}_s(s) ^2} \left(\frac{\partial \tilde{c}_s}{\partial \xi} \right) ^2 + \frac{\kappa (s)}{\tilde{c}_s(s)}\frac{\partial \tilde{c}_s}{\partial \xi} \right. \\
\left.- \frac{1}{2 \tilde{c}_s(s)} \frac{ \partial ^2 \tilde{c}_s}{\partial \xi ^2} \right] \xi = 0~.
\end{split}
\end{equation}
We thus obtain the equation for the deviation between nearby rays as
\begin{equation}
 \frac{1}{\tilde{c}_s(s) } \frac{d }{d s } \left( \frac{1}{\tilde{c}_s(s) }\frac{d \xi}{d s} \right) + K(s) \xi = 0~,
\end{equation}
and this equation is the same as the equation satisfied by $z$ that one can derive from Eqs. (\ref{ham1}-\ref{ham2}).


\section{Eigenvectors and eigenvalues of the monodromy matrix $M$}
\label{alg}
In this section, we derive the necessary formulas to express the stability angle $\alpha$ and the function $\Gamma$ in terms of the elements of the monodromy matrix $M$. 
The characteristic polynomial of a $2\times2$ matrix $M$ is
\begin{equation}
 \lambda ^2 - \mathrm{Tr}(M) \lambda + \mathrm{det}(M) = 0~.
\end{equation}
Then, it is known that a monodromy matrix is a symplectic matrix, which implies that $\mathrm{det}(M)=1$. 
For a stable ray, we have $|\mathrm{Tr}(M)| < 2$, and thus the roots of this polynomial become
\begin{equation}
 \lambda_{1,2} = \frac{\mathrm{Tr}(M)}{2} \pm i \frac{\sqrt{-\mathrm{Tr}(M)^2 + 4 }}{2}~.
\end{equation}
We thus obtain the expression for the stability angle $\alpha$ as
\begin{equation}
\exp(\pm i\alpha) = \frac{1}{2} \left[\mathrm{Tr}(M) \pm i \sqrt{-\mathrm{Tr}(M)^2 + 4} \right]~,
\end{equation}
and we obtain Eq. (\ref{eqalpha}).

The eigenvector $\vec{v}_1$ will then be given by the following equation:
 \begin{equation}
 \begin{pmatrix}
 M_{11}-\lambda_1&M_{12} \\
 M_{21}&M_{22}-\lambda_1
 \end{pmatrix}
\begin{pmatrix}
 z_1 \\
 p_1 
 \end{pmatrix}
 = 0~,
 \end{equation}
which corresponds to
\begin{gather}
 (M_{11}-\lambda_1) z_1 + M_{12} p_1 = 0 \\
 M_{21} z_1 + (M_{22}-\lambda_1) p_1 = 0~.
\end{gather}
Recalling that $\Gamma(s)=p(s)/z(s)$ we write
\begin{equation}
 \Gamma_1 = \frac{p_1}{z_1}~,
\end{equation}
and thus we obtain that
\begin{equation}
 \Gamma_1 = -\frac{(M_{11}-\lambda_1)}{M_{12}} = - \frac{M_{21}}{(M_{22}-\lambda_1)}~.
\end{equation}
This yields
\begin{equation}
 \Gamma_1 = \frac{-M_{11}+M_{22}}{2 M_{12}} + i \frac{\sqrt{- \mathrm{Tr}(M)^2 + 4 }}{2 M_{12}}~,
\end{equation}
and with the same procedure we obtain the expression of $\Gamma_2$:
\begin{equation}
 \Gamma_2 = \frac{-M_{11}+M_{22}}{2 M_{12}} - i \frac{\sqrt{- \mathrm{Tr}(M)^2 + 4 }}{2 M_{12}}~.
\end{equation}
\end{appendix}

\bibliographystyle{aa}
\bibliography{refs}

\end{document}